\pdfoutput=1
 \documentclass[twocolumn,aps,prd,epsfig,nofootinbib]{revtex4}
\usepackage{graphicx,epstopdf,latexsym,amssymb,amsmath,color,mathrsfs,rotating, multirow}
\usepackage[center]{subfigure}

\begin{document}
\allowdisplaybreaks
 \newcommand{\bq}{\begin{equation}} 
 \newcommand{\eq}{\end{equation}}
 \newcommand{\bqn}{\begin{eqnarray}}
 \newcommand{\eqn}{\end{eqnarray}}
 \newcommand{\nb}{\nonumber}
 \newcommand{\lb}{\label}
 \newcommand{\f}{\frac}
 \newcommand{\p}{\partial}
\newcommand{\PRL}{Phys. Rev. Lett.}
\newcommand{\PLB}{Phys. Lett. B}
\newcommand{\PRD}{Phys. Rev. D}
\newcommand{\CQG}{Class. Quantum Grav.}
\newcommand{\JCAP}{J. Cosmol. Astropart. Phys.}
\newcommand{\JHEP}{J. High. Energy. Phys.}
\title{Inflationary spectra with inverse-volume corrections in loop quantum cosmology and their observational constraints from Planck 2015 data}

\author{Tao Zhu${}^{a, b}$}
\email{Tao$\_$Zhu@baylor.edu}

\author{Anzhong Wang $^{a, b}$}
\email{Anzhong$\_$Wang@baylor.edu}

\author{Klaus Kirsten${}^{c}$}
\email{Klaus$\_$Kirsten@baylor.edu}

\author{Gerald Cleaver${}^{d}$}
\email{Gerald$\_$Cleaver@baylor.edu}

\author{Qin Sheng${}^{c}$}
\email{Qin$\_$Sheng@baylor.edu}

\author{Qiang Wu${}^{a}$}
\email{wuq@zjut.edu.cn}

\affiliation{${}^{a}$ Institute for Advanced Physics $\&$ Mathematics, Zhejiang University of Technology, Hangzhou, 310032, China\\
${}^{b}$ GCAP-CASPER, Physics Department, Baylor University, Waco, TX 76798-7316, USA\\
${}^{c}$ GCAP-CASPER, Mathematics Department, Baylor University, Waco, TX 76798-7328, USA\\
${}^{d}$ EUCOS-CASPER, Physics Department, Baylor University, Waco, TX 76798-7316, USA}

\date{\today}

\begin{abstract}

{ We first derive the primordial power spectra,  spectral indices and runnings of both  scalar and tensor perturbations  of a flat inflationary universe  to the second-order approximations of the slow-roll parameters, in the framework of loop quantum cosmology with the inverse-volume  quantum  corrections. This represents an extension of our previous work in which  the parameter $\sigma$ was assumed to be an integer, where $\sigma$ characterizes the quantum corrections and in general can take any of values from the range $\sigma \in (0, 6]$. Restricting to the first-order approximations of the slow-roll parameters, we find corrections to the results obtained previously in the literature, and point out the causes for such errors. To our best knowledge, these represent the most accurate calculations of scalar and tensor perturbations given so far in the literature. Then, fitting the perturbations to the recently released data by Planck (2015),  we obtain the most severe  constraints  for various values of $\sigma$. Using these constraints as our referring point, we discuss whether these quantum gravitational corrections can lead to measurable signatures in the future cosmological observations. We show that, depending on the value of $\sigma$,  the scale-dependent contributions to the relativistic inflationary spectra due to the inverse-volume corrections could be well within the range of the detectability of the forthcoming generations of experiments, such as the Stage IV experiments.}

\end{abstract}


\maketitle

\section{Introduction}
\renewcommand{\theequation}{1.\arabic{equation}} \setcounter{equation}{0}

The quantization of gravity is still an open problem in physics. While there exist various approaches, it is important to calculate their observational predictions or corrections, and then test them. However, quantum gravitational corrections are in general too small to be detected in the near-future man-made terrestrial experiments. As a result, it is difficult to discriminate different quantization schemes. A natural attempt to bridge this gap is to consider the inflationary cosmology, as one believes that the energy scale when inflation starts is usually very close to the Planck energy, and inflation is very sensitive to Planckian physics \cite{DB}.  In general, quantum gravitational effects were encoded in the inflationary fluctuations $\mu_k(\eta)$,   
which evolved during inflation and obeys the modified Mukhanov-Sasaki equation 
\bqn\lb{EoM}
\mu''_k(\eta)+\left(\omega_k^2(\eta)-\frac{z''(\eta)}{z(\eta)}\right)\mu_k(\eta)=0,
\eqn
where $\omega_k^2(\eta)$ is the associated dispersion relation for the inflationary mode function $\mu_k(\eta)$ and $z(\eta)$ depends on the cosmological background evolution and types of perturbations, scalar, vector or tensor. Thus, it is an essential step to extract observational signatures of quantum gravity from the solution of the above equation, in order to test different quantizations of gravity.

Our aim in this paper is to search for the potential observational signatures to the primordial inflationary spectra from Loop Quantum Cosmology (LQC). As mentioned above, such signatures are normally small,  and highly accurate calculations are required. To understand the physics better, analytical calculations are specially desired. Recently, we developed such a technique, the so-called {\em uniform asymptotic approximation} \cite{Zhu1, Zhu2, Uniform3, Uniform4, Zhu3, Olver1974}, which is specially to fulfill this purpose \footnote{This was first applied to inflationary universe in the framework of  general relativity in \cite{uniformPRL} with the dispersion relation $\omega_k^2 = k^2$.}.  In LQC,  two types of quantum corrections are expected, the holonomy \cite{Mielczarek2008,Grain2009PRL,Grain2010PRD,vector_holonomy,scalar,scalar2,Mielczarek2014,loop_corrections}, and inverse-volume corrections \cite{loop_corrections,Bojowald2008,Bojowald2009,Bojowald2007,Bojowald2008b,Bojowald2011,Bojowald2011b}. In the past decades,  lots of efforts have already been devoted to the inflationary predictions of LQC with these two types of quantum corrections. For the holonomy corrections, the anomaly-free cosmological scalar, vector and tensor perturbations are derived in \cite{vector_holonomy, scalar, scalar2}, and then the inflationary predictions for both scalar and tensor (gravitational wave) perturbations were calculated with the slow-roll approximations up to the first-order \cite{Mielczarek2014}. For the inverse-volume corrections, it is shown that the algebras of cosmological scalar \cite{Bojowald2008, Bojowald2009}, vector \cite{Bojowald2007} and tensor perturbations \cite{Bojowald2008b}  can be also closed. The inflationary power spectra of both scalar and tensor perturbations due to the inverse-volume corrections, again with the slow-roll approximation up to the first-order, were studied in \cite{Bojowald2011}, in which some constraints on the parameters of the model were obtained from observational data \cite{Bojowald2011b}.  { It should be noted that in \cite{Bojowald2011}  the horizon crossing
was taken  as $k = {\cal{H}}$. But, due to the quantum gravitational effects, the dispersion relation is modified to the forms  of Eqs.(\ref{scalar-inv}) and (\ref{tensor-inv}), so the horizon crossing should be at $s(\eta) k =  {\cal{H}}$ for the scalar 
perturbations, and at $\alpha(\eta) k =  {\cal{H}}$ for the tensor perturbations. It can be shown that this causes errors even to the first-order approximation of the slow-roll  parameters. 
In addition,   in \cite{Bojowald2011}  the mode function was first obtained at two limits, ${k\gg {\cal{H}}}$ and ${k\ll {\cal{H}}}$, and then
 matched together at the horizon crossing  where $k \simeq  {\cal{H}}$.  This could also lead  to  huge errors \cite{JM}, as neither $\mu_{k\gg {\cal{H}}}$ nor  $\mu_{k\ll {\cal{H}}}$   is a good approximation of the mode function $\mu_{k}$  at the horizon crossing.
In  \cite{Zhu3}, we showed this explicitly    by considering the exact solution  for $\sigma=2$. Therefore, one of the purposes of this paper is to correct the errors caused by these two kinds of approximations adopted  in \cite{Bojowald2011}.}
We also note that the non-Gaussianities with the inverse-volume corrections  \cite{Cai2012} as well as with the holonomy ones \cite{SH} were  studied. In addition,  quantum gravitational effects from LQC  have been also studied by the so-called dressed metric approach \cite{AAN}, and various interesting results were obtained \cite{AB}.

However, with the arrival of the era of precision cosmology and in order to make comparisons and extract tighter constraints on quantum gravitational corrections from the high precision observational data, more accurate calculations of inflationary predictions from specific quantum gravitational models are highly demanded. Recently, using the third-order uniform asymptotic approximation, we derived explicitly the observational quantities of the slow-roll inflation in the framework of LQC with both of the  holonomy and inverse-volume quantum corrections up to the second-order of the slow-roll parameters \cite{Zhu3} \footnote{In these calculations, we assumed that the quantum corrections were small, and the universe already entered its quasi-de Sitter phase, so that the quantum gravitational effects can be well approximated as linear perturbations \cite{Bojowald2011}. However, it is well-known that, due to the quantum gravitational effects, loop quantum cosmology generically leads to a bouncing universe.  The universe usually first experiences a period of super-inflation and then gradually settles down to a slow-roll inflationary phase \cite{AB}. It would be very interesting to study the effects of such a pre-inflationary phase. Later, we shall come back to this issue again.}. These represent the most accurate calculations existing in the literature as far as we know.  But,  with the inverse-volume corrections, in order to get the inflationary spectra analytically we assumed that the parameter $\sigma$ appearing in these corrections  is an integer. In the subsequent numerical simulations with observational data, we obtained the constraints on the inverse-volume corrections, and showed that these quantum corrections could be within the observational range in the forthcoming experiments, if $\sigma$ is about equal to 1.
 
Clearly, it is natural to ask what will happen for the general case with any given value of $\sigma$? In this paper, we shall answer this question, and in particular focus ourselves in the case with  $0 < \sigma \lesssim 1$, as these are also the favorable theoretical values of $\sigma$ \cite{Bojowald2011}. 
With this purpose,  we shall present a new strategy in the uniform asymptotic approximation to derive the spectra and spectral indices of the inflationary universe  with the inverse-volume quantum corrections. 
We shall demonstrate  that this allows us to calculate the power spectra and spectral indices  up to the second-order of the slow-roll approximations for any given  value of $\sigma$. In addition, with the analytical expressions of inflationary predictions we obtain,  and using the recently released Planck 2015 data \cite{Planck2015}, we obtain new constraints on the quantum gravitational effects from the inverse-valume corrections, and find that such effects could still be well within the detection of the forthcoming experiments, specially the State IV cosmological observations \cite{S4-CMB}.

The rest of the paper is arranged as follows. In Sec. II, we present the background evolutions and cosmological perturbation equations in the framework of LQC with the inverse-volume corrections. Then, in Sec III we turn to use the third-order uniform asymptotic approximation to calculate the inflationary power spectra, spectral indices, and running of spectral indices for both scalar and tensor perturbations with the inverse-volume corrections. In Sec IV, we present the analysis of observational constraints on the inverse-volume corrections in details, by fitting our analytical results with the most recently released observational data, the Planck 2015 \cite{Planck2015}.  In doing so, we obtain the tightest constraints on the parameters of LQC with the inverse-volume corrections, existing in the literature so far. However, even with such constraints, we show that there still exists the possibility of detecting these effects in the forthcoming cosmological observations.  These are  summarized in Sec. V. Two appendices are also included, in which more technical calculations are presented.

\section{Background equations and equations of motion for scalar and tensor perturbations}
\renewcommand{\theequation}{2.\arabic{equation}} \setcounter{equation}{0}

Let us  consider a flat FLRW background,
\bqn
ds^2=a^2(\eta)(-d\eta^2+\delta_{ij}dx^idx^j),
\eqn
where $a(\eta)$ is the scale factor of the universe and $\eta$ is the conformal time, which is related to the cosmic time $t$ via the relation $dt = ad\eta$. In the presence of the inverse-volume corrections, the effective Friedmann and Klein-Gordon equations in the above flat FLRW background read \cite{Bojowald2011}, 
\bqn\lb{friedmann}
&& H^2=\frac{8\pi G}{3} \alpha \left(\frac{\dot \varphi^2}{2 \vartheta }+V(\varphi)\right),\\
\lb{KG}
&& \ddot \varphi +H \left(3-2\frac{d\ln \vartheta}{d\ln p}\right) \dot \varphi +\vartheta \frac{dV(\varphi)}{d\varphi}=0,
\eqn
with $p\equiv a^2$ and
\bqn\lb{alpha}
\alpha \simeq 1+\alpha_0 \delta_{\text{Pl}},\;\;\vartheta\simeq 1+\vartheta_0 \delta_{\text{Pl}},
\eqn
where $\delta_{\text{Pl}}$ characterizes the inverse-volume corrections in loop quantum cosmology,
\bqn
\delta_{\text{Pl}} \equiv \left(\frac{a_{\text{Pl}}}{a}\right)^\sigma.
\eqn
In the above expressions, $\alpha_0$, $\vartheta_0$, and $a_{\text{Pl}}$ are constants that depend on the specific models and parametrization of the loop quantization. Specifically for the parameter $\sigma$, different parametrization schemes shall provide different ranges of $\sigma$ \cite{Bojowald2011}.  Moreover, $\alpha_0$ and $\vartheta_0$ are related by the consistency condition
\bqn
\vartheta_0 (\sigma-3)(\sigma+6)=3 \alpha_0 (\sigma-6),
\eqn
while $\sigma$ takes values in the range $0<\sigma\leq 6$.

It should be noted that the above equations are valid only up to the  first-order of $\delta_{\text{Pl}}$  \cite{Bojowald2011}. Therefore, to be consistent, through the whole paper we shall expand all the quantities to its first-order. 
The evolution of the background, which can be determined by the above set of equations, is usually different from the evolution given in the standard slow-roll inflation, because of the purely geometric effects of the inverse-volume corrections. However, as indicated in \cite{Bojowald2011b}, in that regime the constraint algebra has not been shown to be closed. One way to consider the slow-roll inflation with the inverse-volume corrections is to restrict ourselves to the large-volume regime, in which  the quantum corrections are small and the constraint algebra is closed. In this paper, we will focus on this case.

In the framework of the slow-roll inflation with the inverse-volume corrections, there are two independent schemes of approximations, one is the approximation by cutting off all the contributions from terms higher than $\mathcal{O}(\delta_{\text{Pl}})$, while another approximation is based on the expansion of the background evolution in terms of the tiny slow-roll parameters. It is convenient to define the Hubble hierachy parameters as
\bqn
\epsilon_{1}\equiv -\frac{d\ln H}{d\ln a}, \;\;\epsilon_{i+1} \equiv \frac{d\ln \epsilon_{i}}{d\ln a}.
\eqn
Although the slow-roll parameters $\epsilon_i$ and the inverse-volume quantum correction $\delta_{\text{Pl}}$ are based on two independent approximations, in this paper we treat the parameter $\delta_{\text{Pl}}$ as a second-order qunatity of the slow-roll  parameters $\epsilon_i$. This is reasonable, if we look at the observational constraints on $\delta$. In particular,  in Sec. IV we show clearly that the parameter $\delta$ is roughly at the order of $\mathcal{O}(10^{-4})$, which is of order of $\mathcal{O}\left(\epsilon_V^2\right)$. 

The inverse-volume corrections also affect the evolution of cosmological perturbations. In particular, it was found that, when the inverse-volume corrections are present, the gauge-invariant comoving curvature perturbation $\mathcal{R}$ is conserved at large scales.  Such a feature of $\mathcal{R}$ strongly suggests that one can write a simple Mukhanov equation in the variable $\mu_k^{(s)}(\eta) \equiv z_s \mathcal{R}$, which takes the
form \cite{Bojowald2009c, Bojowald2011},
\bqn\lb{scalar-inv}
\frac{d^2\mu_k^{(s)}(\eta)}{d\eta^2}+\left(s^2(\eta) k^2-\frac{z''_s(\eta)}{z_s(\eta)}\right)\mu^{(s)}_k(\eta)=0,
\eqn
where
\bqn
z_s(\eta) \equiv \frac{a\dot \varphi}{H} \left[1+\frac{\alpha_0-2 \vartheta_0}{2} \delta_{\text{Pl}}\right],
\eqn
depends on the evolution of the background,  and
\bqn
s^2(\eta) \equiv 1+\chi \delta_{\text{Pl}},
\eqn
with
\bqn\lb{chi}
\chi \equiv \frac{\sigma \vartheta_0}{3} \left(\frac{\sigma}{6}+1\right)+\frac{\alpha_0}{2} \left(5-\frac{\sigma}{3}\right).
\eqn
In the slow-roll approximation, $z_s''/z_s$ can be casted in the form
\bqn
\frac{z_s''(\eta)}{z_s(\eta)}&\simeq& a^2H^2 \Big[2-\epsilon_1+\frac{3\epsilon_2}{2}-\frac{\epsilon_1\epsilon_2}{2}+\frac{\epsilon_2^2}{4}+\frac{\epsilon_2\epsilon_3}{2}\nb\\
&&~~~~~~~~+f^{(s)}(\epsilon_i) \delta_{\text{Pl}}\Big],
\eqn
where
\bqn
f^{(s)}(\epsilon_i) &\equiv& \frac{\sigma^2 (\sigma-3)\alpha_0}{4\epsilon_1}+\frac{\sigma(\sigma-3)(\sigma+6)\vartheta_0}{12}+\frac{\sigma^2\alpha_0}{4}\nb\\
&&+\frac{\sigma(\sigma-3)\alpha_0}{4}\frac{\epsilon_2}{\epsilon_1}.
\eqn

For the cosmological tensor perturbations $h_k$, when the inverse-volume corrections are present, the corresponding Mukhanov equation for the variable $\mu^{(t)}_k(\eta) \equiv z_t h_k $ takes the form \cite{Bojowald2011},
\bqn\lb{tensor-inv}
\frac{d^2\mu^{(t)}_k(\eta)}{d\eta^2}+\left(\alpha^2(\eta) k^2-\frac{z''_t(\eta)}{z_t(\eta)}\right)\mu^{(t)}_k(\eta)=0,
\eqn
with $\alpha(\eta)$ being given by Eq.\ (\ref{alpha}) and
\bqn
z_t(\eta) \equiv a \left(1-\frac{\alpha_0}{2} \delta_{\text{Pl}}\right).
\eqn
Similarly, in the slow-roll approximation,  one finds
\bqn
\frac{z_t''(\eta)}{z_t(\eta)} \simeq a^2H^2 \Big[2-\epsilon_1+f^{t}(\epsilon_i) \delta_{\text{Pl}}\Big],
\eqn
where
\bqn
f^{(t)}(\epsilon_i) \equiv \frac{\sigma(\sigma-3)}{2}\alpha_0.
\eqn

\section{Inflationary observables in the uniform asymptotic approximation with an arbitrary value of $\sigma$ }
\renewcommand{\theequation}{3.\arabic{equation}} \setcounter{equation}{0}

\subsection{General formula for the inflationary power spectra}

In general, in order to calculate the   cosmological scalar and tensor perturbations, we first write the equation of the mode function, scalar or tensor,  in the form
\bqn\lb{eom}
\frac{d^2\mu_k(y)}{dy^2}=\left\{\lambda^2 \hat g(y)+q(y)\right\}\mu_k(y),
\eqn
where $y=-k\eta$, and
\bqn
\lambda^2 \hat g(y)+q(y) =-\frac{1}{k^2} \left(s^2(\eta)k^2-\frac{z_s''(\eta)}{z_s(\eta)}\right),
\eqn
for the scalar perturbations, and 
\bqn
\lambda^2 \hat g(y)+q(y) =-\frac{1}{k^2} \left(\alpha^2(\eta)k^2-\frac{z_t''(\eta)}{z_t(\eta)}\right),
\eqn
for the tensor perturbations. Since $\delta_{\text{Pl}} \propto a^{-\sigma}$, it is easy to see that $\delta_{\text{Pl}} \propto y^{\sigma}$ in the slow-roll background. With this feature one can parametrize the expression of $\lambda^2\hat g(y)+q(y)$ in the  form
\bqn
\lambda^2 \hat g(y)+q(y)=\frac{\nu^2(\eta)-1/4}{y^2}-1-\chi \delta_{\text{Pl}}+\frac{m(\eta)}{y^2} \delta_{\text{Pl}},\nb
\eqn
and $\delta_{\text{Pl}}$ in the form
\bqn
\delta_{\text{Pl}}=\left(\frac{a_{\text{Pl}}}{a}\right)^\sigma \left(\frac{H}{-a\eta H}\right)^\sigma y^{\sigma}=\kappa(\eta) y^{\sigma} \epsilon_{\text{Pl}},
\eqn
where $\nu(\eta)$, $m(\eta)$, $\nu(\eta)$ are slow-roll quantities depending on the types of the perturbations, $\epsilon_{\text{Pl}} \equiv (a_{\text{Pl}}/k)^{\sigma}\ll 1$ and $\kappa(\eta)\equiv (-a\eta)^{-\sigma}$. Recall 
 that $\chi$ is given by Eq.(\ref{chi}) for the scalar perturbations, while one has to replace $\chi$ by $2\alpha_0$ when considering the tensor perturbations. With the above setup, one finds
\bqn
\lambda^2 \hat g(y)+q(y)&=&\frac{\nu^2(\eta)-1/4}{y^2}-1 - \chi \kappa(\eta) y^\sigma \epsilon_{\text{Pl}}\nb\\
&&~  
+m(\eta)  \kappa(\eta) y^{\sigma-2} \epsilon_{\text{Pl}}.
\eqn 
Note that in this paper, we also use the notation $g(y)=\lambda^2 \hat g(y)$.

In our previous work \cite{Zhu3}, we  applied the uniform asymptotic approximation method to solve the above equations of motion when $\sigma$ is an integer.  However, when the parameter $\sigma$ is any given value, the method becomes not applicable simply because of two reasons. First, when $\sigma$ is not an integer, the functions $\lambda^2 \hat g(y)$ and $q(y)$ may not be analytic functions near their two poles, $y=0^{+}$ and $y=+\infty$, which makes the usual analysis of the convergence for the error control functions given in \cite{Zhu3} invalid. Second,  it becomes extremely difficult to perform the corresponding integral of $\sqrt{\lambda^2 \hat g(y)}$ when $\sigma$ is not an integer. 

In this paper, the first problem is solved by considering a new kind of expansion of $\hat g(y)$ and $q(y)$ near the pole $y\to 0^+$ (Eqs.(\ref{new_expansion})), in which the terms $y^{\sigma}$ are treated,
 separately. By using this kind of expansions, in Appendix A we  prove explcitly that the error control function associated with the approximate solutions near the pole $y = 0$ (as well as the one $y = \infty$) is  convergent, provided that we choose
\bqn
\lambda^2 \hat g(y)&=&\frac{\nu^2}{y^2}-1-\chi \epsilon_{\text{Pl}}\kappa y^{\sigma}+m\epsilon_{\text{Pl}} \kappa y^{\sigma-2},\\
q(y)&=&- \frac{1}{4y^2}.
\eqn
 
With the above choice, the analytical approximate solution of Eq.(\ref{eom}) can be completely determined in the uniform asymptotic approximation,
 and subsequently the general formulas of the power spectra for both the scalar and tensor perturbations can be expressed as
\bqn\lb{power_spectra}
\Delta^2(k)&\equiv& \frac{k^3}{2\pi^2} \left|\frac{\mu_k(y)}{z}\right|^2_{y\to 0^{+}}\nb\\
&\simeq&\frac{k^2}{4\pi^2}\frac{-k\eta}{z^2(\eta) \nu(\eta)}\exp\left(2\lambda  \int_y^{\bar y_0}\sqrt{\hat g(y')}dy'\right)\nb\\
&&\;\times \left[1+\frac{\mathscr{H}(+\infty)}{\lambda}+\frac{\mathscr{H}^2(+\infty)}{2\lambda^2}+\mathcal{O}(1/\lambda^3)\right],\nb\\
\eqn
where $\bar y_0$ denotes the turning point (or zero) of $\hat g(y)$, i.e., $\hat g(\bar y_0)=0$, from which one finds
\bqn
\bar y_0= \nu(\eta_0)+\frac{1}{2}\kappa(\eta_0)[m(\eta_0)-\chi \nu^2(\eta_0)) \nu^{\sigma-1}(\eta_0)]\epsilon_{\text{Pl}},\nb\\
\eqn
with $-\eta_0=\bar y_0/k$. The error control function $\mathscr{H}(+\infty)$  is given by,  
\bqn\lb{H}
\frac{\mathscr{H}(+\infty)}{\lambda}&=&\frac{5}{36}\left.\left\{\int_{\bar y_0}^{y}\sqrt{\lambda^2 \hat g(y')}dy'\right\}^{-1}\right|_{y\to \bar y_0}^{y\to 0^+}\nb\\
&&-\int_{\bar y_0}^{y\to 0^+} \left\{\frac{q}{g}-\frac{5g'^2}{16g^3}+\frac{g''}{4g^2}\right\}\sqrt{g}dy'.\nb\\
\eqn
Thus, in order to calculate the power spectra, one has to perform the integral of $\sqrt{\hat g(y)}$ and the integrals appearing  in Eq.(\ref{H}). In the following, let us consider them separately.

\subsection{Calculations of the integral of $\sqrt{\hat g(y)}$ and the error control function $\mathscr{H}(+\infty)$}

As we mentioned above, when we consider a general $\sigma$, it is extremely difficult to calculate the integral of $\sqrt{g(y)}$ and the integrals appearing in the error control function $\mathscr{H}(\infty)$. In our previous treatment \cite{Zhu3}, in order to calculate these integrals, we expressed the function $\hat g(y)$ in the form
\bqn
g(y)\simeq \frac{\bar y_0-y}{y^2} \Big(h_0+h_1y+\cdots+h_{\sigma+1}y^{\sigma+1}\Big).
\eqn
Note that writing $g(y)$ in the above form plays  an essential role in the calculations of the integrals, and this can only be achieved when $\sigma$ is an integer. 

In order to extend the above calculation to the case for a general value of $\sigma$, in this paper we adopt a new strategy, in which we first expand the integrands in terms of $\delta_{\text{Pl}}$ to its the first-order, and then carry out explicitly the integrals. Only after this, we expand the obtained results in terms of the slow-roll parameters. Previously, we did just in the opposite order. However, cautions must be taken, as such expansions usually do not commute, and different results could be obtained with different orderings, unless proper conditions are imposed. In Appendix B, a concrete example of this kind is given. Fortunately, in our current case we find that such conditions are satisfied, and the finally results are independent of the order of the expansions. 

With the above in mind, 
let us first consider the integral of $\sqrt{g(y)}$. In terms of $\epsilon_{\text{Pl}}$, according to Eq.(\ref{int_of_sqrt_g}), the integral can be expanded as
\bqn
&&\int^{\bar y_0}_{y}\sqrt{g(y')}dy'\nb\\
&&~~\simeq \int^{\bar \nu_0}_{y}\frac{\sqrt{\nu^2-y'^2}}{y'}dy'\nb\\
&&~~~+\epsilon_{\text{Pl}} \int^{\bar \nu_0}_{y} \frac{m \kappa y'^{\sigma-1}-\chi \kappa y'^{\sigma+1}}{2\sqrt{\nu^2-y'^2}}dy'.
\eqn
As we are considering the de Sitter background with the slow-roll approximations, it is convenient to employ the following expansions,
\bqn\lb{sl-ex}
\nu(\eta)&\simeq &\bar \nu_0+\bar \nu_1 \ln\f{y}{\bar \nu_0}+\frac{\bar \nu_2}{2}\ln^2\f{y}{\bar \nu_0},\nb\\
\kappa(\eta)&\simeq& \bar \kappa_0+\bar \kappa_1 \ln\f{y}{\bar \nu_0}+\frac{\bar \kappa_2}{2}\ln^2\f{y}{\bar \nu_0},\nb\\
m(\eta)&\simeq& \bar m_0+\bar m_1 \ln\f{y}{\bar \nu_0}+\frac{\bar m_2}{2}\ln^2\f{y}{\bar \nu_0}.
\eqn
Note that with the above expansions, the turning point $\bar y_0$ can be re-expressed as
\bqn
\bar y_0 \simeq \bar \nu_0+\frac{1}{2}\bar \kappa_0 \Big(\bar m_0-\chi \bar \nu_0^{2}\Big)\bar \nu_0^{\sigma-1}\epsilon_{\text{Pl}},
\eqn
where
\bqn
&&\bar \nu_0 \sim \frac{3}{2}+\mathcal{O}(\epsilon_i),\;\;\bar \nu_1 \sim \mathcal{O}(\epsilon_i^2),\;\;\;\bar \nu_2 \sim \mathcal{O}(\epsilon_i^3),\nb\\
&& \bar \kappa_0 \sim \bar H^\sigma \Big(1+\mathcal{O}(\epsilon_i)\Big),\;\;\bar \kappa_1 \sim \bar H^\sigma \mathcal{O}(\epsilon_i),\;\;\bar \kappa_2 \sim \bar H^\sigma \mathcal{O}(\epsilon_i^2),\nb\\
&& \bar m_0 \sim \mathcal{O}(\epsilon_i^{-1}),\;\;\bar m_1 \sim \mathcal{O}(1),\;\;\bar m_2 \sim  \mathcal{O}(\epsilon_i),
\eqn
here $\epsilon_i$'s represent the slow-roll quantities. As in loop quantum cosmology, the quantity $\delta \equiv \alpha_0 H^\sigma \epsilon_{\text{Pl}} \sim \mathcal{O}(\epsilon_i^2)$, therefore, up to the second-order in the slow-roll expansion, only the terms $\nu_1 $, $m_1 \epsilon_{\text{Pl}}$, $\kappa_1 m_0 \epsilon_{\text{Pl}}$ contribute to the power spectra. Thus,  it is sufficient to consider only these terms in the following calculations. 

Substituting the slow-roll expansions of $\nu(\eta)$, $m(\eta)$, and $\kappa(\eta)$ given  in Eq.(\ref{sl-ex}) into the above integral, and after tedious calculations, we get
\bqn\lb{sl_sqrt_g}
&&\int^{\bar y_0}_{y}\sqrt{g(y')}dy' \nb\\
&&~~~\simeq -\left(1+\ln\frac{y}{2\bar \nu_0}\right)\bar \nu_0\nb\\
&&~~~~~-\left(\f{1}{2}\ln^2\frac{y}{\bar\nu_0}+\frac{\pi^2}{24}-\frac{1}{2}\ln^22 \right)\bar\nu _1 \nb\\
&&~~~~~+\mathscr{G}_1 (\bar m_0+\sigma m_0-\sigma\chi \bar \nu_0^2 ) \bar\kappa_0 \bar \nu_0^{\sigma-1}\epsilon_{\text{Pl}}\nb\\
&&~~~~~+\mathscr{G}_2 \left(\bar \kappa _0 \bar m_1+\bar \kappa _1 \bar m_0\right)\bar \nu _0^{\sigma -1}\epsilon _{\text{Pl}},
\eqn
where $\mathscr{G}_{1, 2}$ are given in Eq.(\ref{C.1}). 
Now, let us
 turn to consider the error control function $\mathscr{H}(+\infty)$ in Eq.(\ref{H}). Substituting the expansions of Eq.(\ref{sl-ex}) into Eq.(\ref{int_exp}), after tedious calculations we finally get
\bqn
\lb{sl_sqrt_ga}
\mathscr{H}(+\infty)&\simeq &\frac{1}{6 \bar \nu _0}-\frac{23+12 \ln2}{72 \bar \nu _0^2}\bar \nu _1\nb\\
&&+\mathscr{G}_3 \bar m_0 \bar \kappa_0\bar \nu_0^{\sigma-3} \epsilon_{\text{Pl}}+\mathscr{G}_4\chi  \bar \kappa_0\bar \nu_0^{\sigma-1} \epsilon_{\text{Pl}}\nb\\
&&+\mathscr{G}_5 \left(\bar m_0 \bar \kappa _1+\bar m_1 \bar \kappa _0\right)  \bar \nu _0^{\sigma-3 }\epsilon_{\text{Pl}},
\eqn
where $\mathscr{G}_{3, 4, 5}$ are given in Eq.(\ref{C.2}).

\subsection{Slow-roll expansion of power spectra for both the scalar and tensor perturbations}

All the slow-roll quantities like $\bar \nu_0$, $\bar \nu_1$, $\bar\kappa_0$, and also $\bar m_0$ and $\bar m_1$,  can be expanded in terms of the slow-roll parameters. Let us first consider $\bar \nu_0$ and $\bar \nu_1$. We have
\bqn
\bar \nu^s_0 &\simeq&\frac{3}{2}+\bar \epsilon _1+\frac{\bar \epsilon _2}{2}+\bar \epsilon _1^2+\frac{11 \bar \epsilon _1\bar \epsilon _2 }{6}+\frac{\bar \epsilon _2 \bar \epsilon _3}{6},\nb\\
\bar \nu^s_1 &\simeq&   -\bar\epsilon _1 \bar\epsilon _2-\frac{\bar\epsilon _3 \bar\epsilon _2}{2},
\eqn
for the scalar perturbations, and
\bqn
\bar \nu_0^t&\simeq &  \frac{3}{2}+\bar \epsilon _1+\bar \epsilon_1^2+\frac{4 \bar \epsilon _1 \bar \epsilon_2}{3}\nb\\
\bar \nu_{1}^t &\simeq & -\bar \epsilon _1 \bar \epsilon _2,
\eqn
for the tensor perturbations. Also by using the Friedmann  and Klein-Gordon equations with the inverse-volume corrections, we find
\bqn
\bar m^s_0 &\simeq& \frac{\sigma^2\left(3- \sigma\right) \alpha _0}{4\bar \epsilon_1}+\left(\frac{3 \sigma }{2}-\frac{\sigma ^2}{4}-\frac{\sigma ^3}{12}\right) \vartheta_0\nb\\
&&+\left(\frac{5 \sigma ^2}{4}-\frac{\sigma ^3}{2}\right) \alpha_0,\\
\bar m^s_1 &\simeq &\frac{\sigma ^2 \alpha _0 (3-\sigma ) \bar \epsilon _2}{4 \bar \epsilon _1},
\eqn
for the scalar perturbations, and
\bqn
\bar m^t_0 &\simeq& \frac{3 \sigma  \alpha _0}{2}-\frac{\sigma ^2 \alpha _0}{2},\\
\bar m_1^t&\simeq &\mathcal{O}(\epsilon_i^2),
\eqn
for the tensor perturbations. For $\bar \kappa_0$, we have
\bqn
\bar \kappa_0 &\simeq&  \bar H^\sigma (1-\sigma  \bar \epsilon _1),\\
\bar \kappa_1 &\simeq & \sigma  \bar H^\sigma \bar \epsilon _1.
\eqn

Now substituting the above slow-roll expansions into the formula of the power spectra, and after tedious calculations,  we find 
\bqn\lb{scalar_spectrum}
\Delta_s^2(k) &\simeq& \bar A_s \Bigg\{1-2 (1+\bar D_{\text{p}}) \bar \epsilon _1-\bar D_{\text{p}} \bar \epsilon _2\nb\\
&&~~~~+\left(2 \bar D_{\text{p}}+2 \bar D_{\text{p}}^2+\frac{\pi ^2}{2}-5+\bar \Delta _1\right) \bar \epsilon _1^2\nb\\
&&~~~~+\left(\bar D_{\text{p}}^2-\bar D_{\text{p}}+\frac{7 \pi ^2}{12}-8+\bar \Delta _1+2 \bar \Delta _2\right) \bar \epsilon _1 \bar \epsilon _2\nb\\
&&~~~~+\left(\frac{1}{2}\bar D_{\text{p}}^2+\frac{\pi ^2}{8}-\frac{3}{2}+\frac{1}{4}\Delta _1\right)\bar \epsilon _2^2\nb\\
&&~~~~+\left(\frac{\pi ^2}{24}+\bar\Delta _2-\frac{1}{2}\bar D_{\text{p}}^2\right) \bar \epsilon _2\bar  \epsilon _3\nb\\
&&~~~~+\epsilon_{\text{Pl}} \left(\frac{3\bar H}{2}\right)^\sigma\left(\frac{\mathcal{\bar Q}^{(s)}_{-1}}{\bar \epsilon_1}+\mathcal{\bar Q}^{(s)}_0+\f{\mathcal{\bar Q}^{(s)}_{1} \bar \epsilon_2}{ \bar \epsilon_1}\right)\Bigg\},\nb\\
\eqn
for the scalar spectrum, where $\bar A_s \equiv \frac{181\bar H^2}{72 e^3\pi^2 \bar \epsilon_1}$, $\bar D_{\text{p}}\equiv \frac{67}{181}-\ln2$, $\bar \Delta_1 \equiv \frac{183606}{32761}-\frac{\pi^2}{2}$, $\bar \Delta_2 \equiv \frac{9269}{589698}$, and
$\mathcal{\bar Q}_{k}^{(s)}$ are given in Eq.(\ref{C.3}).
Note that a letter with an over bar  denotes a quantity evaluated at $y=\bar \nu^s_0$. It is worthwhile to point out that here the coefficients, $\mathcal{\bar Q}_{-1}^{(s)}, \mathcal{\bar Q}_{0}^{(s)}$, and $\mathcal{\bar Q}_{1}^{(s)}$ look a little bit different from those given in \cite{Zhu3}, by simply taking $\sigma$ to be integers. This is because  the scalar spectrum in the current paper is evaluated at a different point. Here we evaluate the scalar spectrum at the point $y=\bar \nu_0$, while in \cite{Zhu3} it was evaluated at the turning point $y=\bar y_0 \simeq \bar \nu_0 +\frac{1}{2}\kappa_0 (\bar m_0-\chi \bar \nu_0^2) \bar \nu_0^{\sigma-1} \epsilon_{\text{Pl}}$. Actually, as we shall show in the next subsection, when $\sigma$ is an integer, the scalar spectra given in this paper shall reduce precisely to the ones obtained in \cite{Zhu3}, after we express them shift them all at the point when the scalar mode crosses the Hubble horizon. 

Similarly, for the tensor perturbations, we find
\bqn
\Delta_t^2(k) &\simeq& \bar A_t \Bigg\{1-2\left(1+\bar D_{\text{p}}\right) \bar \epsilon _1\nb\\
&&~~~~+\left(\bar \Delta _1+\frac{\pi ^2}{2}-5+2 \bar D_{\text{p}}+2 \bar D_{\text{p}}^2\right) \bar \epsilon _1^2\nb\\
&&~~~~+\left(2 \bar \Delta _2-2+\frac{\pi ^2}{12}-2 \bar D_{\text{p}}-\bar D_{\text{p}}^2\right) \bar \epsilon _1 \bar \epsilon _2\nb\\
&&~~~~+\epsilon_{\text{Pl}} \left(\frac{3}{2}\bar H\right)^\sigma \mathcal{\bar Q}^{(t)}_0\Bigg\},
\eqn
where
\bqn
\mathcal{\bar Q}_0^{(t)}&=&-\frac{2\sigma  \left(\sigma ^2-2 \sigma +6\right)}{3}\alpha _0 \mathscr{G}_1 -\frac{80\sigma(\sigma -3)  }{543} \alpha _0 \mathscr{G}_3 \nb\\
&&+\frac{240 }{181}\alpha _0 \mathscr{G}_4.
\eqn

\subsection{Expansion at Horizon Crossing}

Note that in the above section, all the quantities in the expressions of the power spectra were evaluated at the time $y=-k\bar \eta= \nu^s(\bar \eta)$ (here $\bar \nu_0=\nu^s(\bar\eta)$) for the scalar spectrum, and $y=-k\bar \eta= \nu^t(\bar \eta)$ (here $\bar \nu^s_0=\nu^s(\bar\eta)$) for the tensor spectrum. However, in the conventional treatments, all the observables are usually expanded at the time when the inflationary scalar or tensor modes across the Hubble horizon. Since in general for the same wavenumber $k$, the scalar  and tensor modes cross the horizon at different times, it is reasonable to rewrite the expressions of the scalar and tensor perturbations in terms of  their own time of  horizon-crossing. However, detailed analysis shows the differences between the two different evaluation times only produce high-order corrections in the slow-roll approximation, and can be safely neglected. Thus, in this paper,  we will not distinguish these two different times and only consider the expansions at the time when the scalar mode crosses the Hubble horizon. 

With the above in mind, let us first consider the scalar spectrum. In order to rewrite all expressions in terms of quantities evaluated at horizon crossing, one has to transfer the quantities which are evaluated at the time $y=-k\bar \eta=\nu^s(\bar \eta)$ to the ones evaluated at the time when the scalar mode crosses the Hubble horizon $s(\eta_\star)k=a(\eta_\star) H(\eta_\star)$, where
\bq
s(\eta_\star)=1+\frac{1}{2}\chi \epsilon_{\text{Pl}} \kappa(\eta_\star) (-k\eta_\star)^\sigma.
\eq
This can be achieved by using the expansion
\bqn
f(\bar\eta) =f(\eta_\star)+f_{\star 1} \ln\frac{\bar \eta}{\eta_\star}+\frac{f_{\star 2}}{2} \ln^2\frac{\bar \eta}{\eta_\star}+\cdots,
\eqn
where $\ln\frac{\bar \eta}{\eta_\star}$ can be replaced by
\bqn
\ln\frac{\bar \eta}{\eta_\star}=\ln\left(\frac{\nu_0^s(\eta_\star) (1+\frac{1}{2}\epsilon_{\text{Pl}}H_\star^\sigma)}{a_\star H_\star \eta_\star}\right).
\eqn
Using the above relations, the scalar spectrum of Eq.(\ref{scalar_spectrum}) can be rewritten in the form
\bqn
\Delta_s^2(k) &\simeq &A_s^{\star}\Bigg\{ 1-2\left(1+D_{\text{p}}^{\star}\right) \epsilon _{\text{$\star $1}}-D_{\text{p}}^{\star} \epsilon _{\text{$\star $2}}\nb\\
&&~~+\left(2D_{\text{p}}^{\star 2}+2D_{\text{p}}^{\star}+\frac{\pi^2}{2}-5+\Delta_1^{\star}\right) \epsilon _{\text{$\star $1}}^2\nb\\
&&~~+\left(\frac{1}{2}D_{\text{p}}^{\star 2}+\frac{\pi^2}{8}-1+\frac{\Delta_1^{\star}}{4}\right) \epsilon _{\text{$\star $2}}^2\nb\\
&&~~+\left(D_{\text{p}}^{\star 2}-D_{\text{p}}^{\star}+\frac{7\pi ^2}{12}-7+\Delta_1^{\star}+2\Delta_2^{\star}\right) \epsilon _{\text{$\star $1}} \epsilon _{\text{$\star $2}}\nb\\
&&~~+\left(\frac{\pi ^2}{24}-\frac{1}{2}D_{\text{p}}^{\star 2}+\Delta_2^{\star}\right) \epsilon _{\text{$\star $2}} \epsilon _{\text{$\star $3}}\nb\\
 &&~~+\epsilon_{\text{Pl}} \left(\frac{3H_\star }{2}\right)^\sigma \Big[\frac{\mathcal{Q}_{-1}^{\star (s)}}{\epsilon_{\star 1}}+\mathcal{Q}^{\star (s)}_{0}+\frac{\mathcal{Q}^{\star (s)}_1\epsilon_{\star 2}}{\epsilon_{\star 1}}\Big]\Bigg\},\nb\\
\eqn
where the subscript ``$\star$" denotes evaluation at the horizon crossing, $A^{\star}_{s}\equiv \frac{181 H_\star^{2}}{72 e^3 \pi^2 \epsilon_{\star 1}}$, $D_{\text{p}}^{\star}=\frac{67}{181}-\ln3$, $\Delta_1^{\star} =\frac{485296}{98283}-\frac{\pi^2}{2}$, $\Delta_2^{\star} =\frac{9269}{589698}$, and
\bqn
\mathcal{Q}^{\star (s)}_{-1}&=& \mathcal{\bar Q}^{(s)}_{-1},\nb\\
\mathcal{Q}^{\star (s)}_{0}&=&(\sigma +2) \mathcal{\bar Q}^{(s)}_{-1}\ln\frac{3}{2}+\mathcal{\bar Q}^{(s)}_0,\nb\\
\mathcal{Q}^{\star(s)}_1&=&2\mathcal{\bar Q}^{(s)}_{-1}\ln\frac{3}{2}+\mathcal{\bar Q}^{(s)}_1.
\eqn
The corresponding spectral index for the scalar perturbations can be calculated from the scalar spectrum via the definition
\bqn
n_s\equiv1+\frac{d\ln\Delta_s^2(k)}{d\ln k}.
\eqn
It is important to note that the time $\eta_\star$ when the scalar mode across the Hubble horizon is a function of $k$, since $s(\eta_\star)k=a(\eta_\star) H(\eta_\star)$. Using this relation, one finds
\bqn
\eta_\star \simeq \frac{a(\eta_\star) \eta_\star H(\eta_\star) }{k} \left(1-\frac{1}{2}\chi \epsilon_{\text{Pl}} H^\sigma (\eta_\star)\right).
\eqn
Recall that
\bqn
a(\eta_\star)\eta_\star H(\eta_\star) \simeq -1-\epsilon_{\star1}-\epsilon_{\star1}^2-\epsilon_{\star 1} \epsilon_{\star 2},
\eqn
from which  one finds
\bqn\lb{nsh}
n_s &\simeq &1-2 \epsilon _{\star1}-\epsilon _{\star2} -2 \epsilon _{\star1}^2-  (2 D^{\star}_\text{p}+3) \epsilon _{\star 1} \epsilon _{\star 2}-D^\star_\text{p} \epsilon _{\star2} \epsilon _{\star3}\nb\\
&&+\epsilon _{\text{Pl}}\left(\frac{3H_\star}{2}\right)^{\sigma }  \left(\frac{\mathcal{K}^{\star (s)}_{-1}}{\epsilon _{\star1}}+\mathcal{K}^{\star (s)}_0+\frac{\mathcal{K}^{\star (s)}_1 \epsilon _{\star2}}{\epsilon _{\star1}}\right),\nb\\
\eqn
where
\bqn
\mathcal{K}^{\star (s)}_{-1}&=&-\sigma \mathcal{Q}^{\star (s)}_{-1},\nb\\
\mathcal{K}^{\star (s)}_{0}&=&-\sigma (2D^\star_{\text{p}}+3)\mathcal{Q}^{\star (s)}_{-1}-\sigma\mathcal{Q}^{\star (s)}_{0},\nb\\
\mathcal{K}^{\star(s)}_{1}&=&-(\sigma D^\star_{\text{p}}+1)\mathcal{Q}^{\star (s)}_{-1}-\sigma\mathcal{Q}^{\star (s)}_{1}.
\eqn
Similarly,  the running of the scalar spectral index can be expressed in the form
\bqn\lb{ash}
\alpha_s&\simeq& -2 \epsilon _{\star1} \epsilon _{\star2}-\epsilon _{\star2} \epsilon _{\star3}\nb\\
&&+\epsilon _{\text{Pl}}\left(\frac{3H_\star}{2}\right)^{\sigma } \left(\frac{\mathcal{L}^{\star (s)}_{-1}}{\epsilon _{\star1}}+\frac{\mathcal{L}^{\star (s)}_1 \epsilon _{\star2}}{\epsilon _{\star1}}+\mathcal{L}^{\star (s)}_0\right),\nb\\
\eqn
where
\bqn
\mathcal{L}^{\star (s)}_{-1}&=&-\sigma \mathcal{K}^{\star (s)}_{-1},\nb\\
\mathcal{L}^{\star (s)}_{0}&=&-\sigma \mathcal{K}^{\star (s)}_{-1}-\sigma\mathcal{K}^{\star (s)}_{0},\nb\\
\mathcal{L}^{\star(s)}_{1}&=&-\mathcal{K}^{\star (s)}_{-1}-\sigma\mathcal{K}^{\star (s)}_{1}.
\eqn
It is important to note that, since the inverse-volume corrections can also provide significant contributions to the running spectral indices, it is crucial to consider the terms higher than the second-order runnings. These higher order runnings can be calculated via $\alpha_s^{(l)}\equiv \frac{d^{l}\ln\Delta_s^2(k)}{d\ln^l k}$ (note that $\alpha_s=\alpha_s^{(2)}$ in this notation), from which one finds
\bqn\lb{aslh}
\alpha_s^{(l)}&\simeq& (-1)^{l-1} \sigma^{l-1} \epsilon_{\text{Pl}}\left(\frac{3H_{\star}}{2}\right)^\sigma\nb\\
&&\times \Bigg\{\frac{\mathcal{K}^{\star(s)}_{-1}}{\epsilon_{\star1}}+(l-1)\mathcal{K}^{\star(s)}_{-1}+\mathcal{K}^{\star(s)}_{0}\nb\\
&&~~+\left(\frac{l-1}{\sigma} \mathcal{K}^{\star(s)}_{-1}+\mathcal{K}^{\star(s)}_{1}\right)\frac{\epsilon_{\star2}}{\epsilon_{\star1}}\Bigg\}.
\eqn

Similarly, for the tensor spectrum one finds
\bqn
\Delta_t^2(k) & \simeq & 
A_t^\star \Bigg\{1-2\left(1+D_{\text{p}}^{\star}\right) \epsilon _{\text{$\star $1}}\nb\\
&&~~~+\left(2D_{\text{p}}^{\star 2}+2D_{\text{p}}^{\star}+\frac{\pi^2}{2}-5+\Delta_1^{\star}\right) \epsilon _{\text{$\star $1}}^2\nb\\
&&~~~+\left(-D_{\text{p}}^{\star 2}-2D_{\text{p}}^{\star}+\frac{\pi ^2}{12}-2+2\Delta_2^{\star}\right) \epsilon _{\text{$\star $1}} \epsilon _{\text{$\star $2}}\nb\\
&&~~~+\epsilon_{\text{Pl}} \left(\frac{3 H_\star}{2}\right)^\sigma\mathcal{Q}_0^{\star (t)}\Bigg\},
\eqn
where $\mathcal{Q}_0^{\star (t)}=\mathcal{\bar Q}_0^{(t)}$. The tensor spectral index $n_t$, which is defined via $n_t\equiv d\ln{\Delta_t^2(k)}/d\ln k$, is given by
\bqn\lb{nth}
n_t &\simeq &-2 \epsilon _1^2-2(D^\star_\text{p}+1) \epsilon _{\star1} \epsilon _{\star2}+\epsilon _{\text{Pl}} \left(\frac{3H_\star}{2}\right)^{\sigma } \mathcal{K}^{\star (t)}_0,\nb\\
\eqn
with $\mathcal{K}^{\star (t)}_0=-\sigma \mathcal{Q}^{\star (t)}_0$. Then, the running of the tensor spectral index can be cast in the form,
\bqn\lb{ath}
\alpha_t &\simeq &-2 \epsilon _{\star1} \epsilon _{\star2}+  \epsilon _{\text{Pl}}\left(\frac{3H_\star}{2}\right)^{\sigma }  \mathcal{L}^{\star (t)}_0,
\eqn
and high-order runnings of the spectral index read
\bqn\lb{atlh}
\alpha_t^{(l)}\simeq (-1)^{l-1}\sigma^{l-1} \epsilon_{\text{Pl}}\left(\frac{3H_\star}{2}\right)^{\sigma}\mathcal{K}^{\star (t)}_0.
\eqn
Finally, with both the scalar and tensor spectra given above, one obtains the tensor-to-scalar ratio,
\bqn
r=16 \epsilon_{\star1} \left\{1+D^\star_{\text{p}}\epsilon_{\star2}-\epsilon_{\text{Pl}}\left(\frac{3H_\star}{2}\right)\frac{\mathcal{Q}^{\star (s)}_{-1}}{\epsilon_{\star1}} \right\}.
\eqn

As a consistent  check of the above results, one can compare all the above expressions with those given in \cite{Zhu3}, in which we employed a different approach and the results are  valid only when $\sigma$ is an integer. It
can be shown that the results presented above  reduces exactly to the ones given in \cite{Zhu3},  when we take $\sigma$ to be an integer! This is certainly expected but not obvious, specially after such tedious and 
complicated calculations.  

\section{Observational constraints on inverse-volume corrections}
\renewcommand{\theequation}{4.\arabic{equation}} \setcounter{equation}{0}

In this section, we consider observational constraints on the inverse-volume corrections by carrying out the CMB likelihood analysis for the scalar spectrum we derived in the above section. For this purpose, we run the Cosmological Monte Carlo (CosmoMC) code \cite{COSMOMC} with the recently released data of Planck 2015 \cite{Planck2015}, by assuming the $\Lambda$CDM model.

Since the inverse-volume corrections produce non-negligible contributions to the runnings in both of the scalar and tensor spectra, to proceed,  let us first expand them to all orders in the perturbation wavenumber about a pivot scale $k_\star$. For the scalar spectrum, this expansion reads
\bqn
\ln\Delta_s^2(k)&=&\ln\Delta_s^2(k_0)+\left[n_s(k_0)-1\right]\ln\frac{k}{k_0}+\frac{\alpha_s(k_0)}{2}\ln^2\frac{k}{k_0}\nb\\
&&+\sum_{l=3}^{+\infty} \frac{\alpha_s^{(l)}(k_0)}{l!}\ln^{l}\frac{k}{k_0},
\eqn
where $n_s(k_0)$, $\alpha_s(k_0)$, and $\alpha_s^{(l)}(k_0)$ are given by Eqs. (\ref{nsh}), (\ref{ash}), and (\ref{aslh}), respectively. Similar to \cite{Bojowald2011b}, it is easy to find
\bqn
&&\sum_{l=3}^{+\infty} \frac{\alpha_s^{(l)}(k_0)}{l!}\ln^{l}\frac{k}{k_0}\nb\\
&&~~~~~~~~=\epsilon_{\text{Pl}}\left(\frac{3H_\star}{2}\right)^{\sigma} \nb\\
&&~~~~~~~~~~\times \Bigg\{\mathcal{C}_{1}\frac{\mathcal{K}^{\star(s)}_{-1}}{\epsilon_{\star1}}+\mathcal{C}_1\mathcal{K}^{\star(s)}_0+\mathcal{C}_2\mathcal{K}^{\star(s)}_{-1}\nb\\
&&~~~~~~~~~~~~~~~+\left(\mathcal{C}_1\mathcal{K}^{\star(s)}_1+\mathcal{C}_2\frac{\mathcal{K}^{\star(s)}_{-1}}{\sigma}\right)\frac{\epsilon_{\star2}}{\epsilon_{\star1}}\Bigg\},
\eqn
where
\bqn
\mathcal{C}_1&=&-\ln\frac{k}{k_0}+\frac{\sigma}{2}\ln^2\f{k}{k_0}+\frac{1-e^{-\sigma \ln\frac{k}{k_0}}}{\sigma},\nb\\
\mathcal{C}_2&=&e^{-\sigma \ln\frac{k}{k_0}}\ln\frac{k}{k_0}+\frac{\sigma}{2}\ln^2\f{k}{k_0}+\frac{e^{-\sigma \ln\frac{k}{k_0}}-1}{\sigma}.\nb\\
\eqn

Similarly,  for the tensor spectrum, one finds
\bqn
\ln\Delta_t^2(k)&=&\ln\Delta_t^2(k_0)+n_t(k_0)\ln\frac{k}{k_0}+\frac{\alpha_t(k_0)}{2}\ln^2\frac{k}{k_0}\nb\\
&&+\sum_{l=3}^{+\infty} \frac{\alpha_t^{(l)}(k_0)}{l!}\ln^{l}\frac{k}{k_0},
\eqn
with
\bqn
\sum_{l=3}^{+\infty} \frac{\alpha_t^{(l)}(k_0)}{l!}\ln^{l}\frac{k}{k_0}=\epsilon_{\text{Pl}}\left(\frac{3H_\star}{2}\right)^{\sigma}\mathcal{C}_1\mathcal{K}^{\star(t)}_0,
\eqn
where $n_t(k_0)$, $\alpha_t(k_0)$, and $\alpha_t^{(l)}(k_0)$ are given by Eqs. (\ref{nth}), (\ref{ath}), and (\ref{atlh}), respectively.

In order to carry out the CMB likelihood analysis, it is convenient to introduce the following potential slow-roll parameters,
\bqn\lb{potential_slow_roll}
\epsilon_V\equiv \frac{M_{\text{Pl}}^2}{2}\frac{V_\varphi^2}{V^2},\;\;\eta_V\equiv \frac{M_{\text{Pl}}^2V_{\varphi\varphi}}{V},\;\;\xi_V^2\equiv \frac{M_{\text{Pl}}^4 V_{\varphi}V_{\varphi\varphi\varphi}}{V^2},\nb\\
\eqn
with which we have
\bqn
n_s&\simeq& 1-6\epsilon_V+2\eta_V-\left(24D_{\text{p}}+\frac{10}{3}\right)\epsilon_V^2\nb\\
&&+\left(16D_{\text{p}}-2\right)\epsilon_V \eta_V+\frac{2}{3}\eta_V^2+\left(\frac{2}{3}-2D_{\text{p}}\right)\xi_V^2\nb\\
&&+\frac{\epsilon_{\text{Pl}}H_\star^\sigma}{\epsilon_V} \left\{\frac{3^\sigma}{2^\sigma}\mathcal{K}_{-1}^{\star (s)}+\frac{\sigma^2(\sigma-3)\alpha_0}{18}(3D_{\text{p}} \sigma-\sigma-3)\right\},\nb\\
\eqn
and
\bqn
\alpha_s&\simeq& -24 \epsilon_V^2+16\epsilon_V \eta_V-2\xi_V^2\nb\\
&&+\frac{\epsilon_{\text{Pl}}H_\star^\sigma}{\epsilon_V}\left(\frac{3^\sigma}{2^\sigma}\mathcal{L}_{-1}^{\star (s)}+\frac{\sigma^3(\sigma-3)\alpha_0}{6}\right).
\eqn
For the tensor-to-scalar ratio, we also find
\bqn
r &\simeq& 16\epsilon_V \Bigg\{1+\left(4 D_\text{p}-\frac{4}{3}\right) \epsilon _V+\left(\frac{2}{3}-2 D_\text{p}\right) \eta _V\nb\\
&&+\frac{\epsilon_{\text{Pl}}H^\sigma_\star}{\epsilon_V}\left[-\frac{3^\sigma}{2^\sigma}\mathcal{Q}_{-1}^{\star (s)}+\frac{\sigma(\sigma^2-9)(D_{\text{p}}\sigma-1)\alpha_0}{18}\right]\Bigg\}.\nb\\
\eqn

\subsection{Power-law potential}

We first consider the power-law potential
\bqn
V(\phi)=V_0 \phi^{n}.
\eqn
Here $V_0$ and $n$ are constants. Using the definitions of the potential slow-roll parameters of Eq.(\ref{potential_slow_roll}), we have
\bqn
&&\epsilon_V=\frac{M_{\text{Pl}}^2}{2}\frac{n^2}{\phi^2},\;\;\;\eta_V=M_{\text{Pl}}^2 \frac{n(n-1)}{\phi^2},\nb\\
&&\xi^2_V =M_{\text{Pl}}^4 \frac{n^2(n-1)(n-2)}{\phi^4},
\eqn
from which we have the relations
\bqn
\eta_V=\frac{2(n-1)}{n} \epsilon_V,\;\;\xi_V^2=\frac{4(n-1)(n-2)}{n^2} \epsilon_V^2.
\eqn
In this paper, we also parametrize the inverse-volume corrections by
\bqn
\delta(k)=\alpha_0 \epsilon_{\text{Pl}} H_\star^\sigma.
\eqn
Thus, for the scalar spectrum with the power-law potential, there are only two independent  parameters $\epsilon_V(k_0)$ and $\delta(k_0)$.

To place observational constraints on both $\epsilon_V$ and $\delta$, we run the Cosmological Monte Carlo (CosmoMC) code \cite{COSMOMC} with the latest Planck 2015 data \cite{Planck2015}. For the power-law potential, we focus our attention on $n=1, \frac{2}{3}$, $\frac{1}{3}$. In each of these cases, we choose $\sigma$ as  $\sigma=0.9, 0.8, 0.7, 0.6, 0.5, 0.4$, respectively. In addition, we adopt the flat cold dark matter model with the effective number of neutrinos $N_{\mbox{eff}}=3.046$ and fix the total neutrino mass  as $\sum m_\nu=0.06 \mbox{eV}$. We vary the seven parameters: (i) baryon density parameter, $\Omega_b h^2$, (ii) dark matter density parameter, $\Omega_c h^2$, (iii) the ratio of the sound horizon to angular diameter, $\theta$, (iv) the reionization optical depth $\tau$, (v) $\delta/\epsilon_V$, (vi) $\epsilon_V$, and $\Delta_s^2(k)$. We take the pivot scale $k_0=0.05 \mbox{Mpc}^{-1}$ and $k_0=0.002\mbox{Mpc}^{-1}$, respectively.

The constraints on $\epsilon_V$ and $\delta/\epsilon_V$ from Planck 2015 data are all given in Table I for $k_0=0.05 \mbox{Mpc}^{-1}$ and in Table II for $k_0=0.002\;\mbox{Mpc}^{-1}$, respectively. The two-dimensional marginalized joint confidence contours for $(\delta/\epsilon_V, \epsilon_V)$ are illustrated in Fig.1 for $k_0=0.05 \mbox{Mpc}^{-1}$ and in Fig.2 for $k_0=0.002 \mbox{Mpc}^{-1}$. In both figures, we only display the cases for $\sigma=0.9, \; 0.8, \;0.6$ and $0.4$. 

Let us first consider the results for $k_0=0.05 \mbox{Mpc}^{-1}$. For the same value of $\sigma$, a decrease of $n$ (the index of power-law potential) leads to a decrease of the upper bounds of the parameter $\epsilon_V$, while the upper bounds on $\frac{\delta}{\epsilon_V}$ stay almost the same. A decrease of $n$ also leads to a decrease of the upper bounds of $\delta$, while in order to derive the bounds of $\delta$ we have used the best-fit values of $\epsilon_V$.  For the same potential (with fixed value of $n$), a decrease of $\sigma$ leads to an increase of the upper bounds of $\delta/\epsilon_V$, while it only leads to a slight change on the bounds of $\epsilon_V$. However, as $\delta/\epsilon_V$ is increasing, the upper bound on $\delta$ is also increasing when the value of $\sigma$ is decreasing. {\em The smaller values of $\sigma$, the less tight constraints on $\delta$ are.} Note that the constraints on $\delta/\epsilon_V$ were derived in \cite{Zhu3} for $\sigma=1$ and $\sigma=2$, which are tighter than the ones presented here but for   values of $\sigma<1$. This is expected, and is the main reason why in the current paper we only consider the case with $\sigma < 1$. Since for $\sigma > 2$, it is expected that the constraints become so strong, and the corresponding quantum gravitational effects are extremely small, so it is impossible to detect such effects with the forthcoming generation of experiments, as shown explicitly in \cite{Zhu3}.   

Since the scalar power spectrum with the inverse-volume corrections is $k$-dependent, one expects that changing the pivot scale shall lead to changes to the constraints. For this purpose, we further consider the constraints on $\delta/\epsilon_V$ and $\epsilon_V$ by taking the pivot scale $k_0=0.002 \mbox{Mpc}^{-1}$, for which  all the results are illustrated in Table II and Fig. 2. As it is expected, changing the pivot scale from $k_0=0.05\;\mbox{Mpc}^{-1}$ to $k_0=0.002\;\mbox{Mpc}^{-1}$ indeed  leads to the increase of the upper bounds of $\delta/\epsilon_V$, and consequently increases the upper bound of $\delta$.

\begin{figure*}
\subfigure[$\;\;\sigma=0.9, n=1$]{\label{sigma09-1}
\includegraphics[width=4cm]{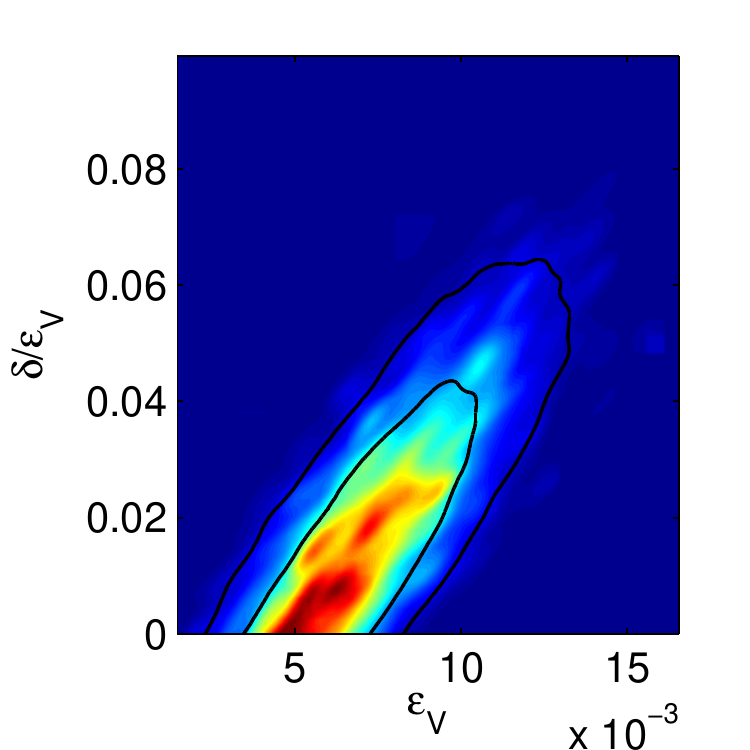}}
\subfigure[$\;\;\sigma=0.9, n=\frac{2}{3}$]{\label{sigma09-2}
\includegraphics[width=4cm]{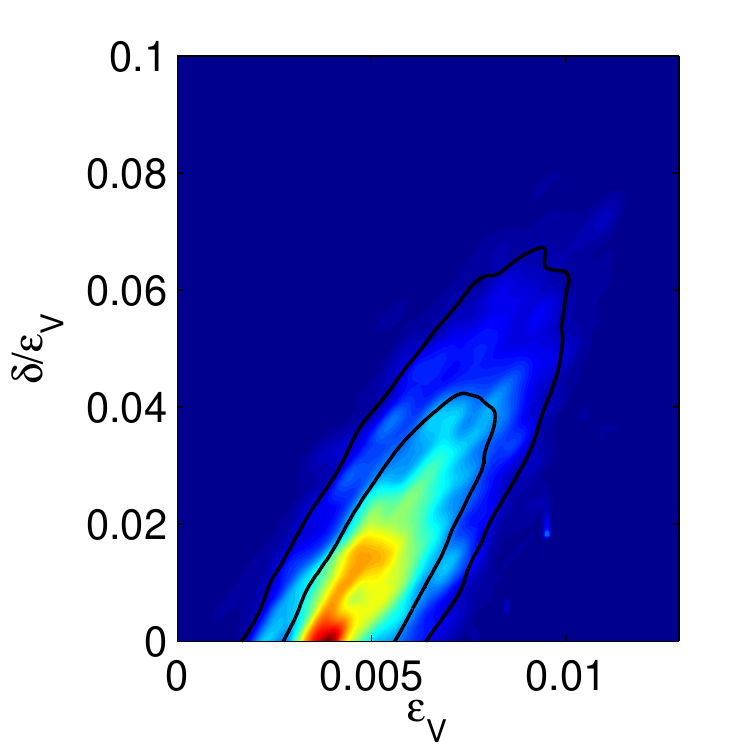}}
\subfigure[$\;\;\sigma=0.9, n=\frac{1}{3}$]{\label{sigma09-3}
\includegraphics[width=4cm]{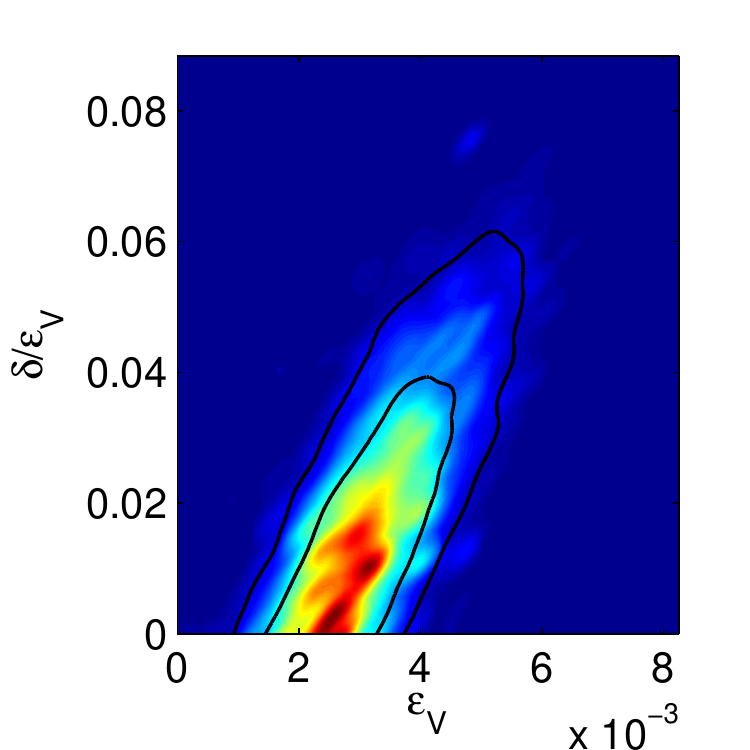}}
\subfigure[$\;\;\sigma=0.8, n=1$]{\label{sigma08-1}
\includegraphics[width=4cm]{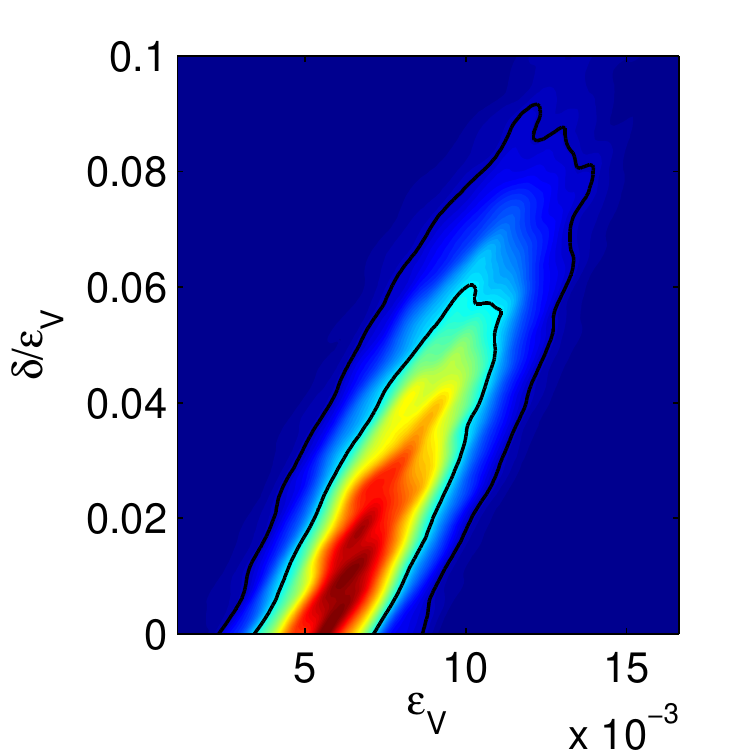}}
\\
\subfigure[$\;\;\sigma=0.8, n=\frac{2}{3}$]{\label{sigma08-2}
\includegraphics[width=4cm]{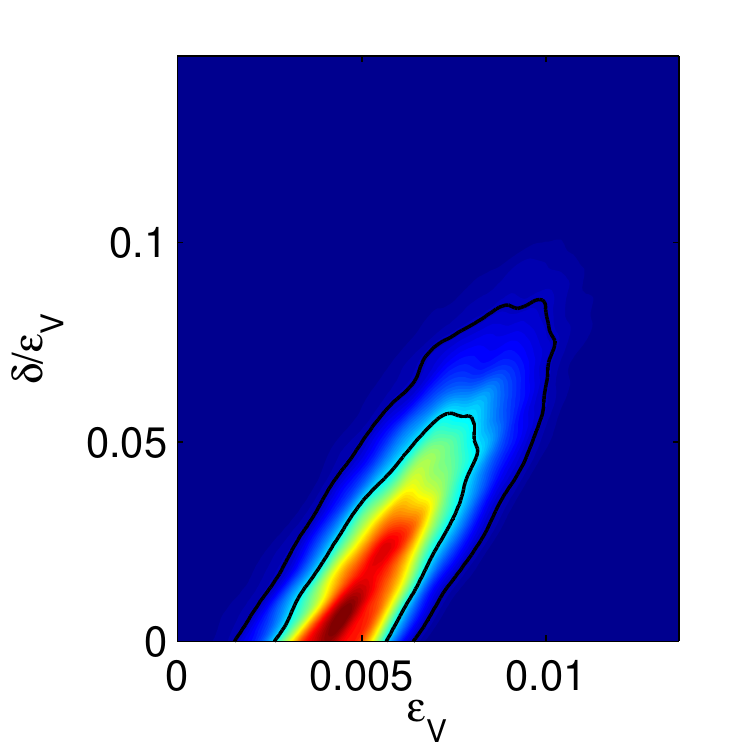}}
\subfigure[$\;\;\sigma=0.8, n=\frac{1}{3}$]{\label{sigma08-3}
\includegraphics[width=4cm]{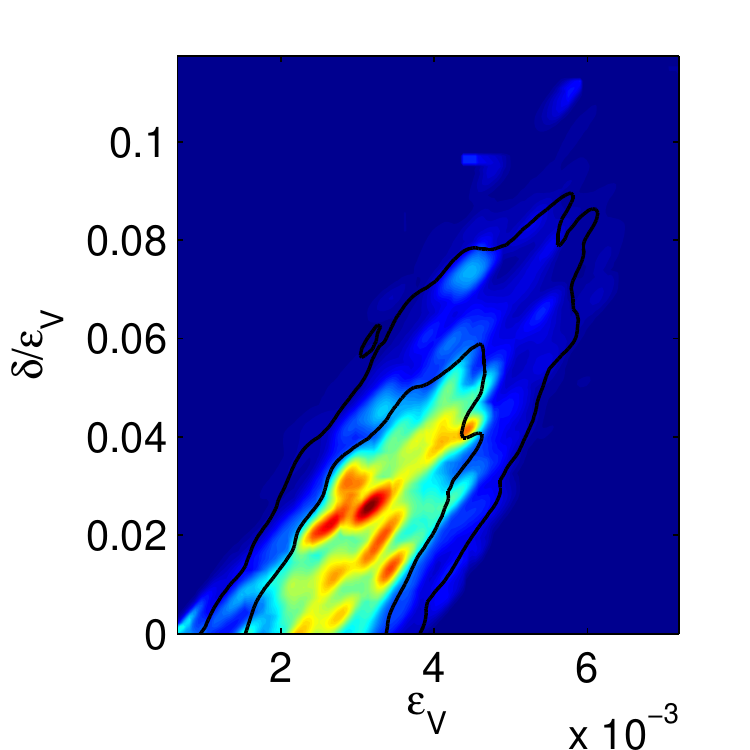}}
\subfigure[$\;\;\sigma=0.6, n=1$]{\label{sigma06-1}
\includegraphics[width=4cm]{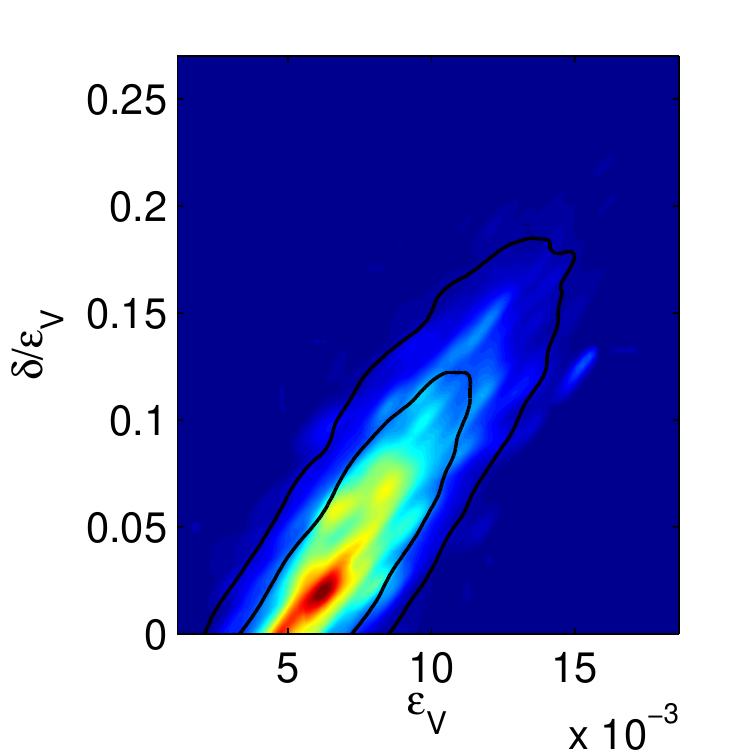}}
\subfigure[$\;\;\sigma=0.6, n=\frac{2}{3}$]{\label{sigma06-2}
\includegraphics[width=4cm]{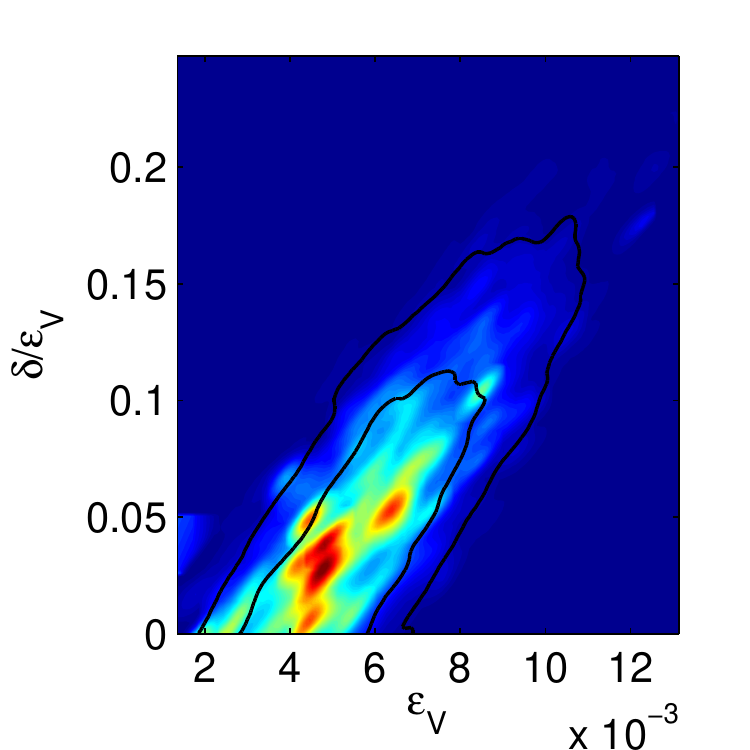}}
\\
\subfigure[$\;\;\sigma=0.6, n=\frac{1}{3}$]{\label{sigma06-3}
\includegraphics[width=4cm]{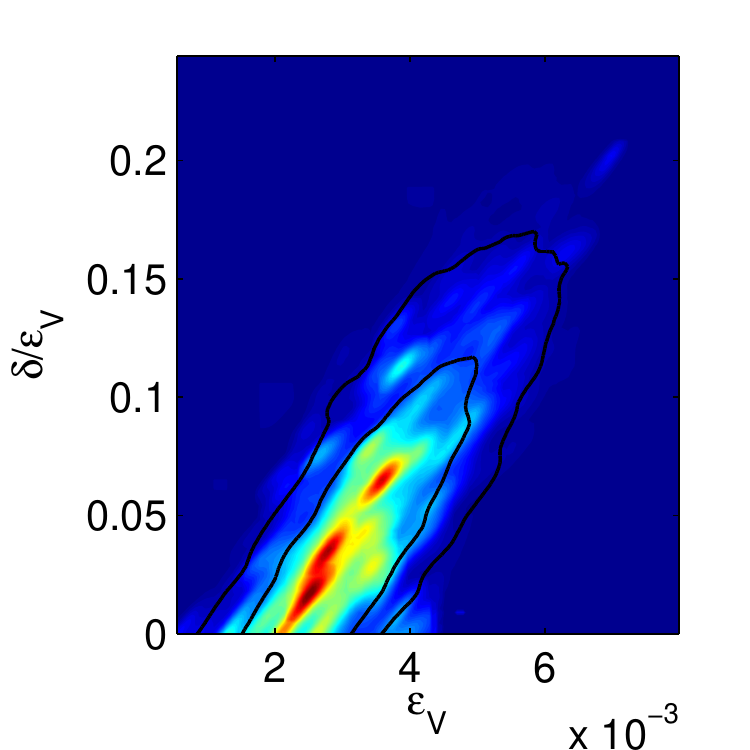}}
\subfigure[$\;\;\sigma=0.4, n=1$]{\label{sigma04-1}
\includegraphics[width=4cm]{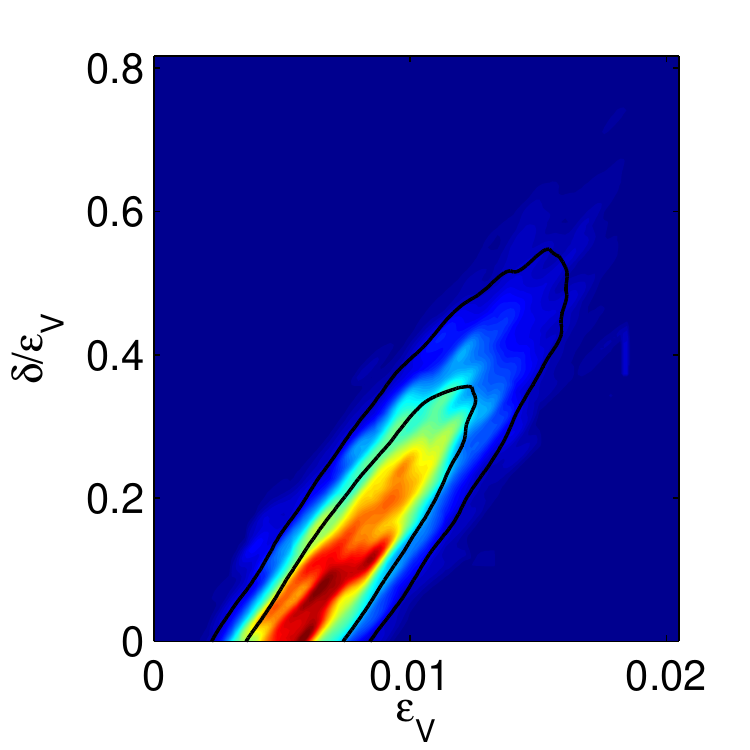}}
\subfigure[$\;\;\sigma=0.4, n=\frac{2}{3}$]{\label{sigma04-2}
\includegraphics[width=4cm]{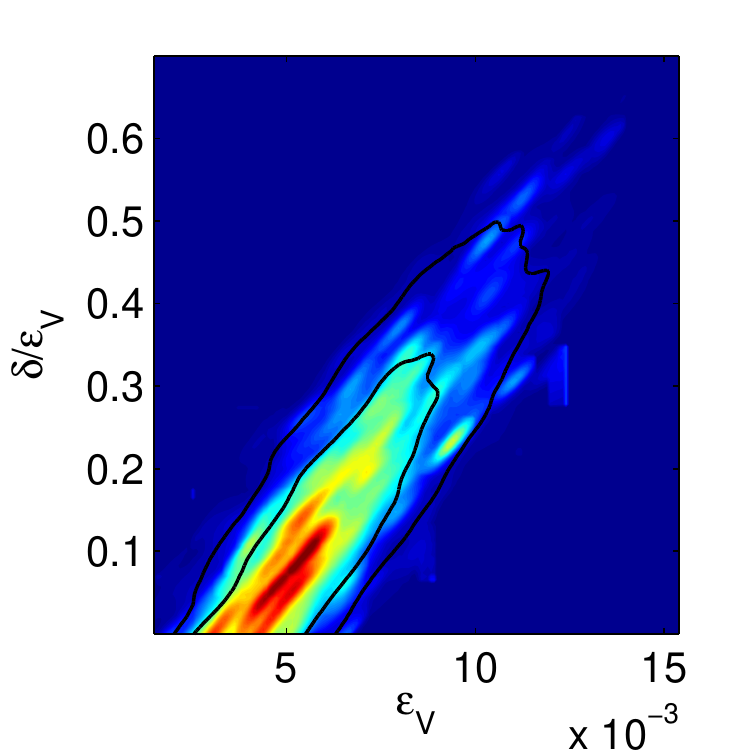}}
\subfigure[$\;\;\sigma=0.4, n=\frac{1}{3}$]{\label{sigma04-3}
\includegraphics[width=4cm]{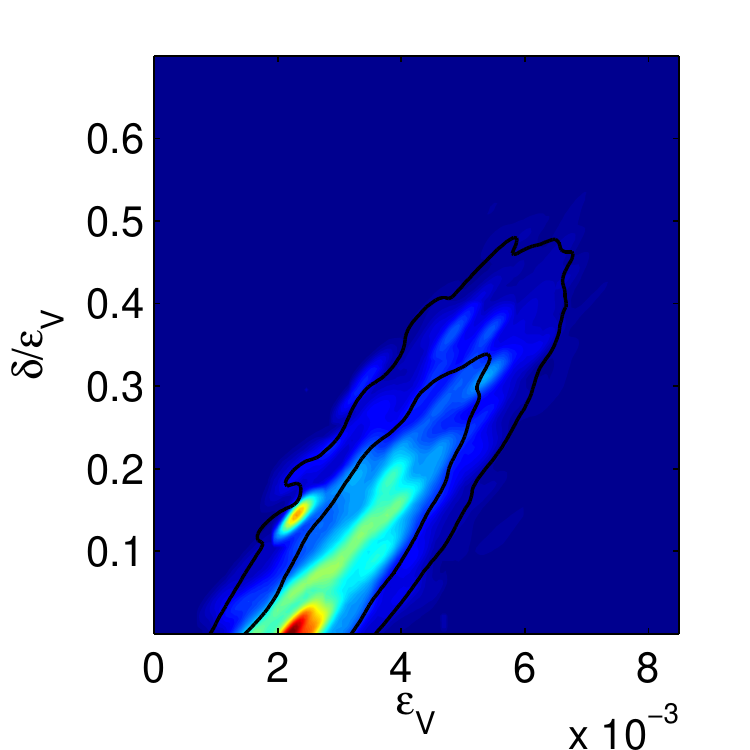}}
\caption{Two-dimensional marginalized distribution for the parameters $\delta/\epsilon_V$ and  $\epsilon_V$ at the pivot scale $k_0=0.05\text{Mpc}^{-1}$ for the power-law potential with $n=1$, $n=\frac{2}{3}$, and $n=\frac{1}{3}$, respectively. The internal and external lines correspond to the confidence levels of  $68\%$ and $95\%$, respectively.} \label{fig1}
\end{figure*}

\begin{figure*}
\subfigure[$\;\;\sigma=0.9, n=1$]{\label{sigma09-4}
\includegraphics[width=4cm]{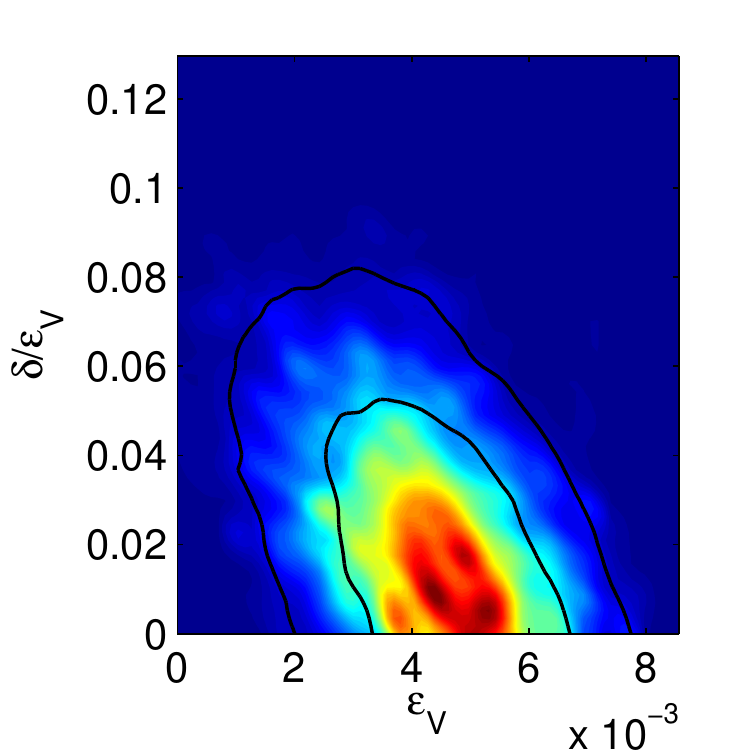}}
\subfigure[$\;\;\sigma=0.9, n=\frac{2}{3}$]{\label{sigma09-5}
\includegraphics[width=4cm]{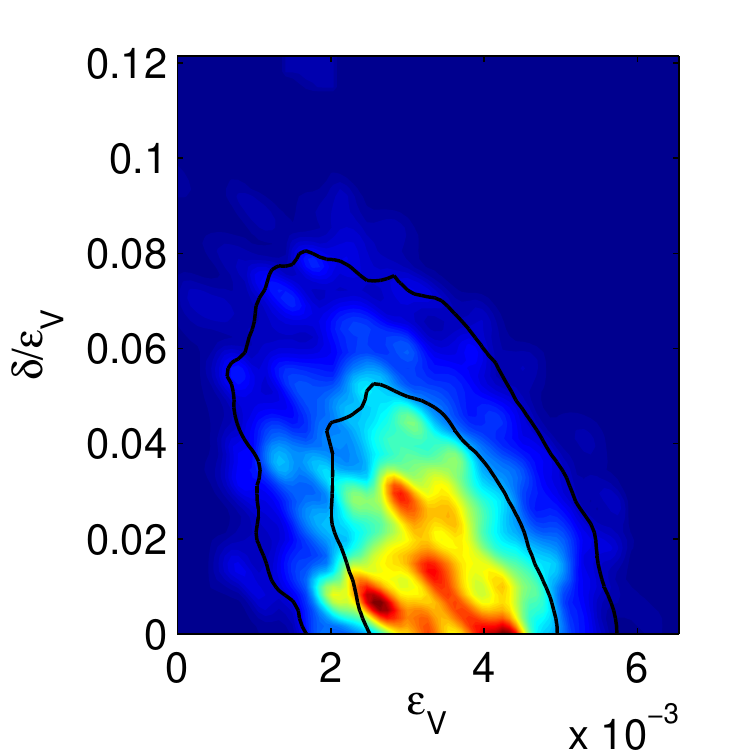}}
\subfigure[$\;\;\sigma=0.9, n=\frac{1}{3}$]{\label{sigma09-6}
\includegraphics[width=4cm]{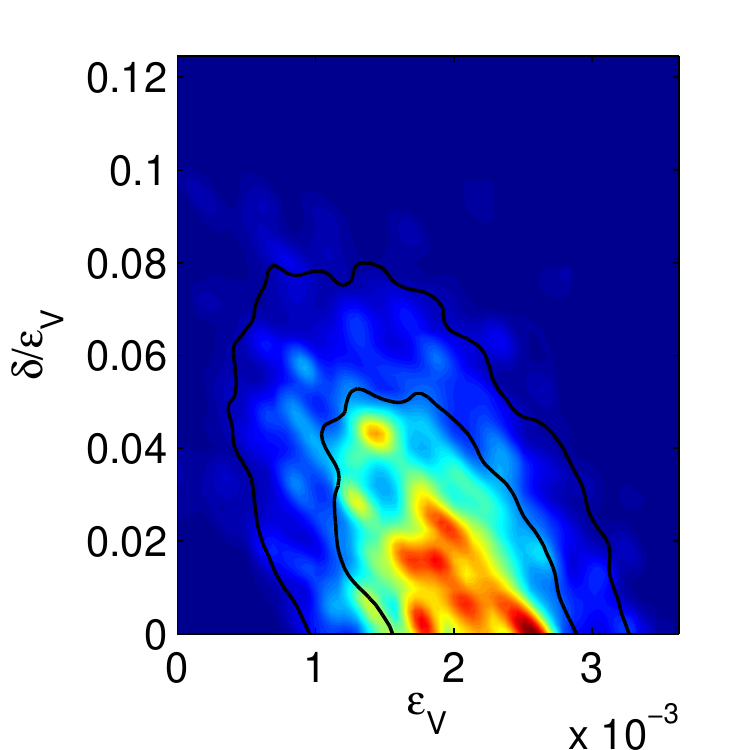}}
\subfigure[$\;\;\sigma=0.8, n=1$]{\label{sigma08-4}
\includegraphics[width=4cm]{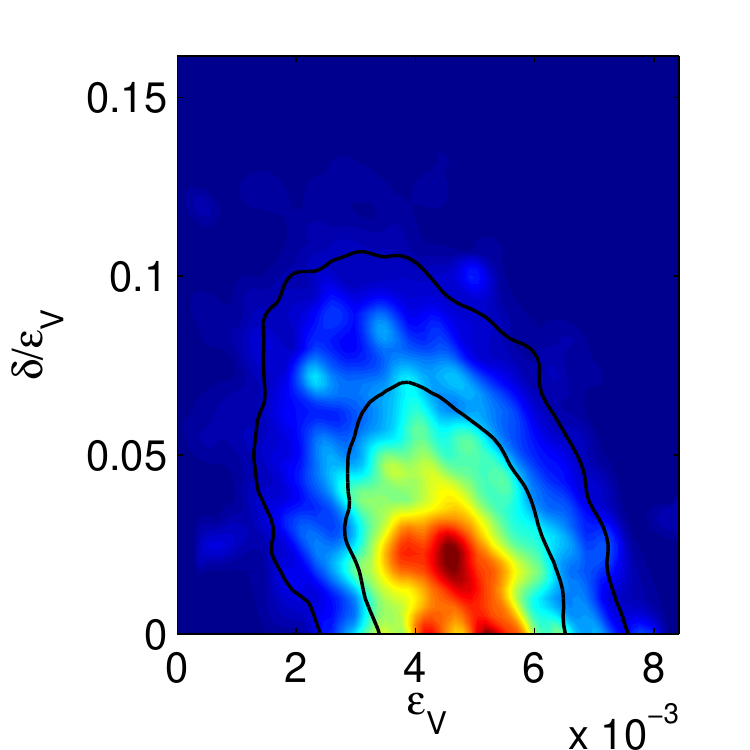}}
\\
\subfigure[$\;\;\sigma=0.8, n=\frac{2}{3}$]{\label{sigma08-5}
\includegraphics[width=4cm]{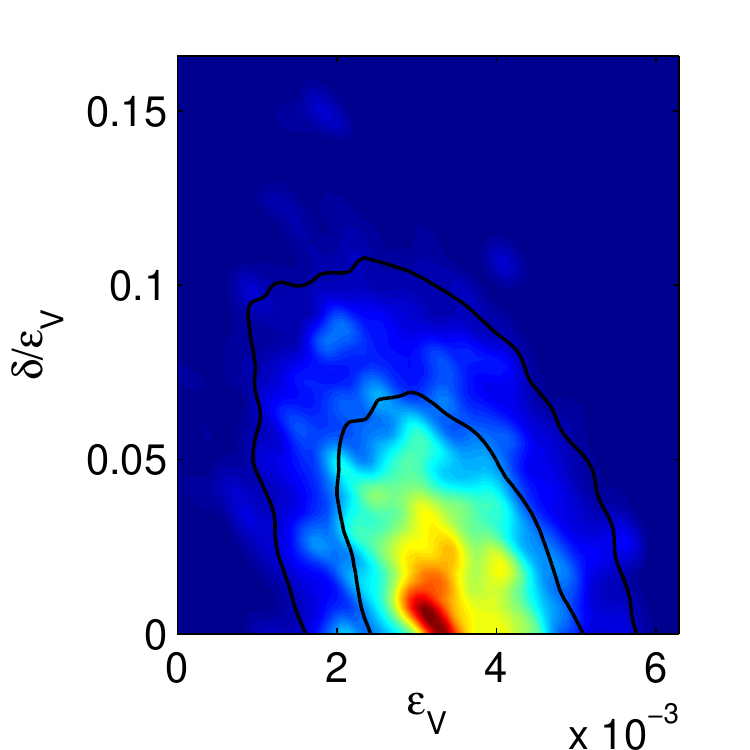}}
\subfigure[$\;\;\sigma=0.8, n=\frac{1}{3}$]{\label{sigma08-6}
\includegraphics[width=4cm]{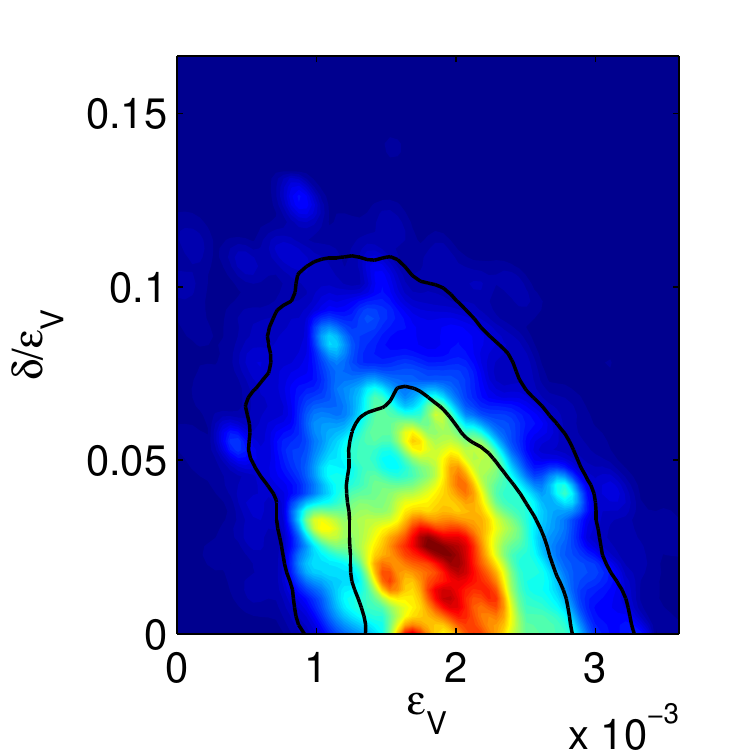}}
\subfigure[$\;\;\sigma=0.6, n=1$]{\label{sigma06-4}
\includegraphics[width=4cm]{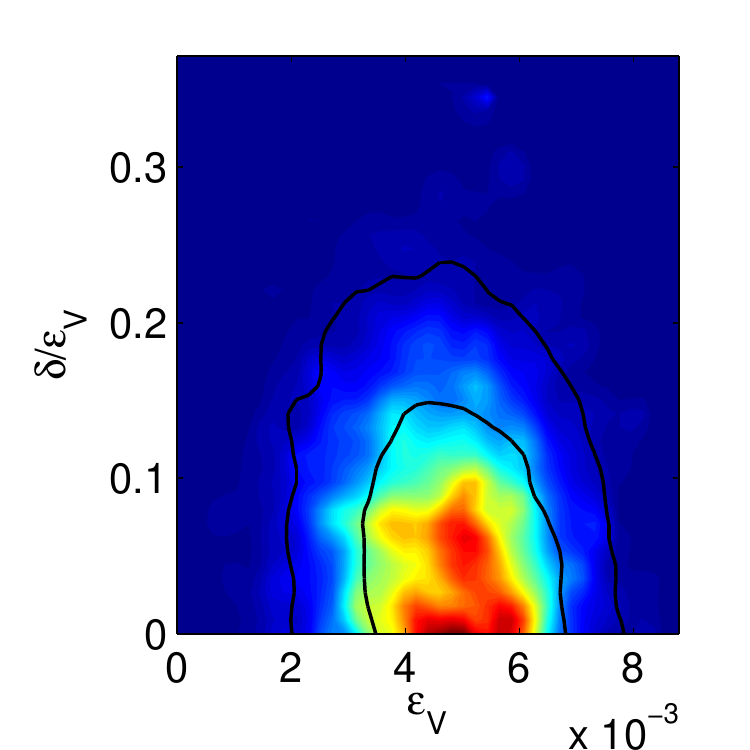}}
\subfigure[$\;\;\sigma=0.6, n=\frac{2}{3}$]{\label{sigma06-5}
\includegraphics[width=4cm]{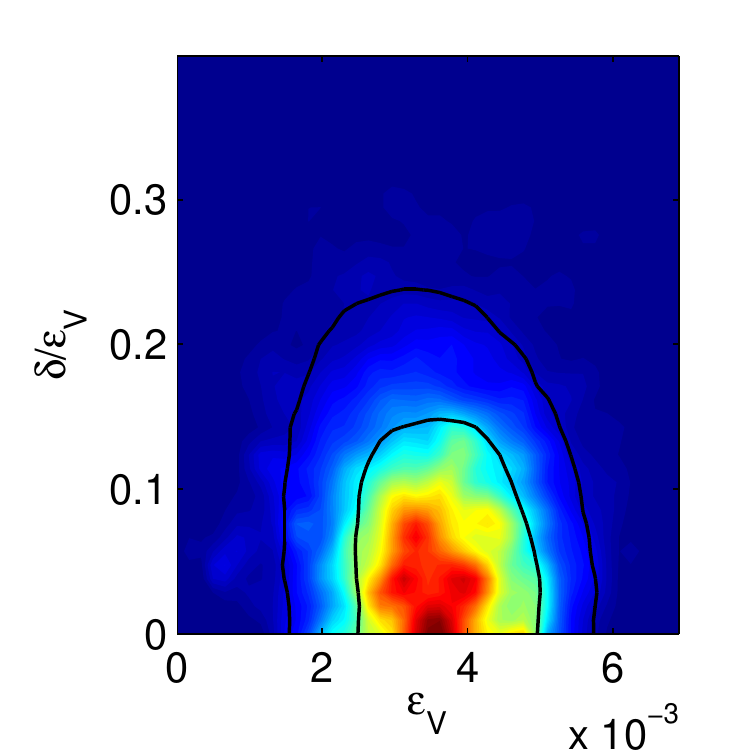}}
\\
\subfigure[$\;\;\sigma=0.6, n=\frac{1}{3}$]{\label{sigma06-6}
\includegraphics[width=4cm]{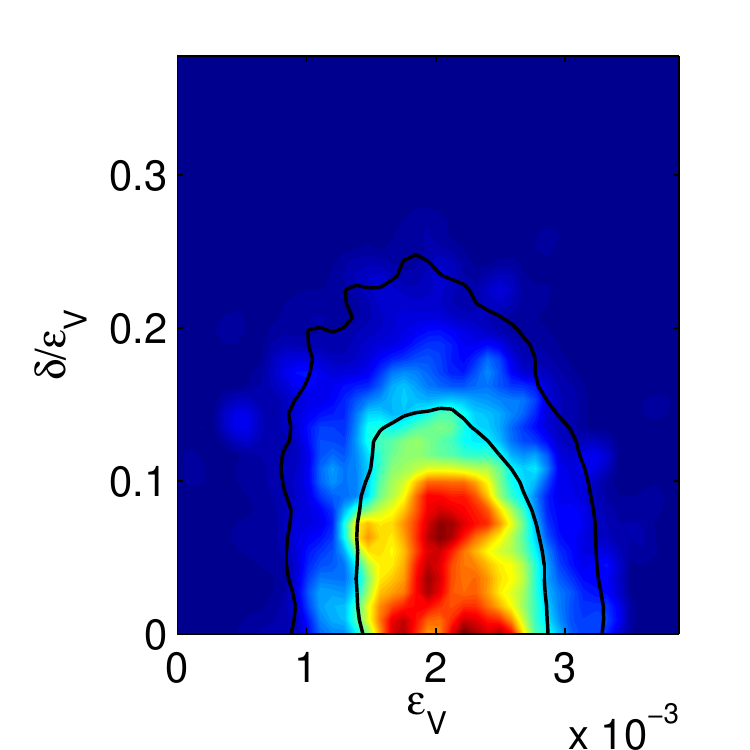}}
\subfigure[$\;\;\sigma=0.4, n=1$]{\label{sigma04-4}
\includegraphics[width=4cm]{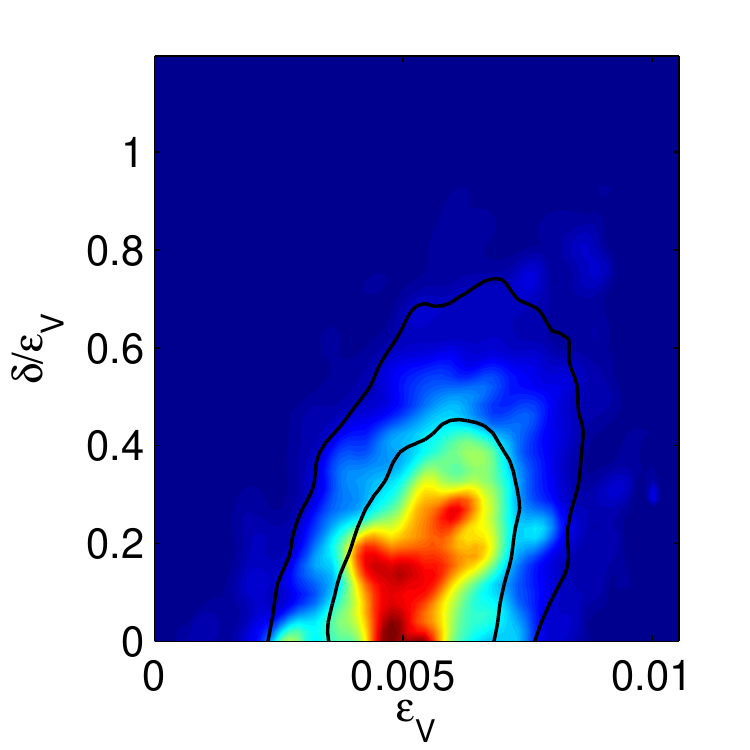}}
\subfigure[$\;\;\sigma=0.4, n=\frac{2}{3}$]{\label{sigma04-5}
\includegraphics[width=4cm]{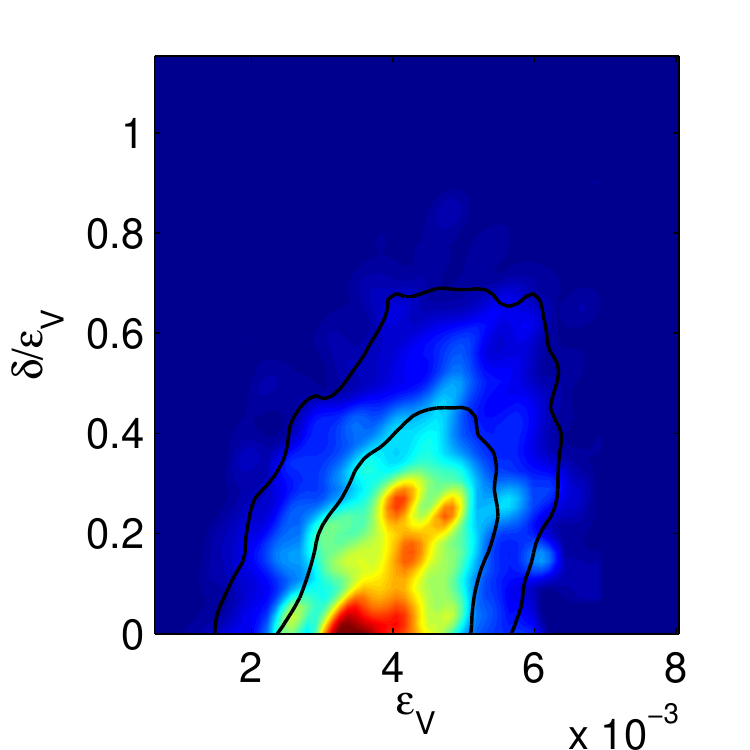}}
\subfigure[$\;\;\sigma=0.4, n=\frac{1}{3}$]{\label{sigma04-6}
\includegraphics[width=4cm]{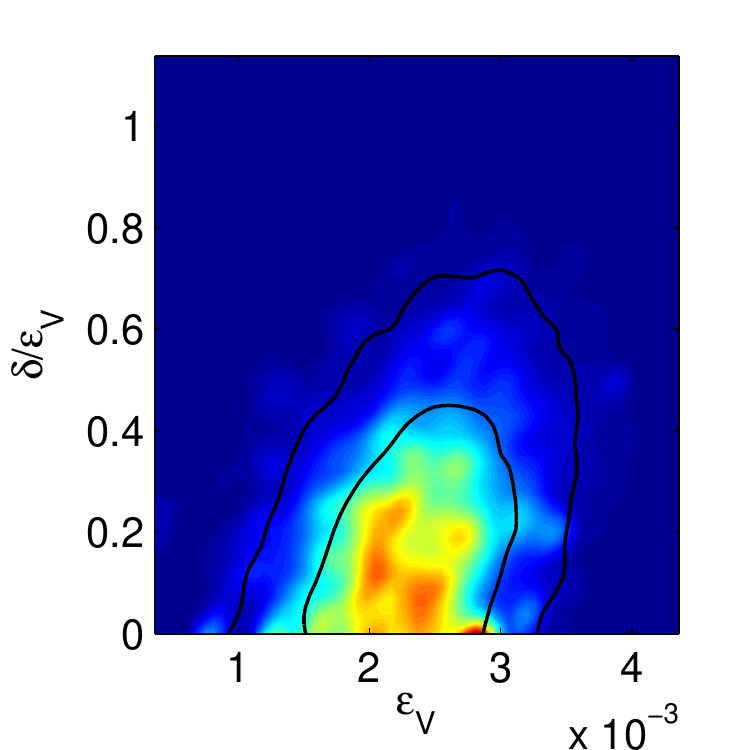}}
\caption{Two-dimensional marginalized distribution for the parameters $\delta/\epsilon_V$ and  $\epsilon_V$ at the pivot scale $k_0=0.002\text{Mpc}^{-1}$ for the power-law potential with $n=1$, $n=\frac{2}{3}$, and $n=\frac{1}{3}$, respectively. The internal and external lines correspond to the confidence levels of  $68\%$ and $95\%$, respectively.} \label{fig2}
\end{figure*}

\subsection{Starobinsky Potential}

The Starobinsky inflationary  model is a realization of inflation based on modified gravity with the action \cite{Starobinsky},
\bqn
S= \int d^4 x \sqrt{-g} \frac{M_{\text{Pl}}^2}{2} \left(R+\frac{R^2}{6M^2}\right),
\eqn
where $M$ is a free parameter of the theory \footnote{Recently, the same potential was studied in the dressed metric approach of LQC \cite{BG}.}. This model can be conformally transferred into Einstein's gravity with a scalar field that has a potential of the form,
\bqn
V(\phi)=\frac{3}{4}M^{2}M_{\text{Pl}}^2\left(1-e^{-\sqrt{2/3}\phi/M_{\text{Pl}}}\right)^2,
\eqn 
 with which the slow-roll parameters have the relations
\bqn
\eta_V&=&\epsilon_V-\frac{2\sqrt{3}}{3} \sqrt{\epsilon_V},\nb\\
\xi_V^2 &=& \frac{4}{3}\epsilon_V-2 \sqrt{3} \epsilon_V^{3/2}.
\eqn
Similar to the last subsection, here in the scalar spectrum we only have two independent parameters, $\epsilon_V(k_0)$ and $\delta(k_0)$. 

Repeating the CMB likelihood analysis given above,
constraints on $\delta/\epsilon_V$ and $\epsilon_V$ are summarized in Table I and Table II for $k_0=0.05 \mbox{Mpc}^{-1}$ and $k_0=0.002 \mbox{Mpc}^{-1}$, respectively. The two-dimensional probability distributions are illustrated in Fig. 3 for both pivot scales at $k_0=0.05 \; \mbox{Mpc}^{-1}$ and $k_0=0.002 \; \mbox{Mpc}^{-1}$. Comparing the constraints with that obtained for power-law potential, it is interesting to see that the upper bound on $\epsilon_V$ dramaticly decreases, while the upper bound on $\delta/\epsilon_V$ increases. As a result, we also see that the upper bound of $\delta$ becomes tighter than those in the power-law potential case for the same value of $\sigma$. Unlike in power-law potential case where the upper bounds of $\delta/\epsilon_V$ become less tight when we change the pivot scale $k_0=0.05\; \mbox{Mpc}^{-1}$ to $k_0=0.05 \; \mbox{Mpc}^{-1}$, here the bounds are stronger at the pivot scale $k_0=0.002\;\mbox{Mpc}^{-1}$ for the same values of $\sigma$. 

{
From Tables I and II, one can see that  the upper bound of $\delta/\epsilon_V$   is,  
\bqn
\lb{UBs}
\frac{\delta}{\epsilon_V} \lesssim  \mathcal{O}(10^{-2}) \sim \mathcal{O}(10^{-1}),
\eqn
depending on the specific value of $\sigma$.  In addition, this upper bound is not sensitive to the forms of the inflationary potential $V(\phi)$. In particular, for a given $\sigma$ it is almost the same for different values of the power-law index  $n$. These same bounds are also applicable to the Starobinsky potential, as long as the parameter $\sigma$ is the same. That is, the upper bound (\ref{UBs}) is quite independent of  inflationary models. 
This is very important when we consider the quantum gravitational effects in the next subsection, as the conclusion regarding to the detectability obtained from (\ref{UBs}) is expected to be model-independent, too.   
}

\begin{figure*}
\subfigure[$\;\;\sigma=0.9, k_0=0.05\rm{Mpc}^{-1}$]{\label{sigma09-st-1}
\includegraphics[width=4cm]{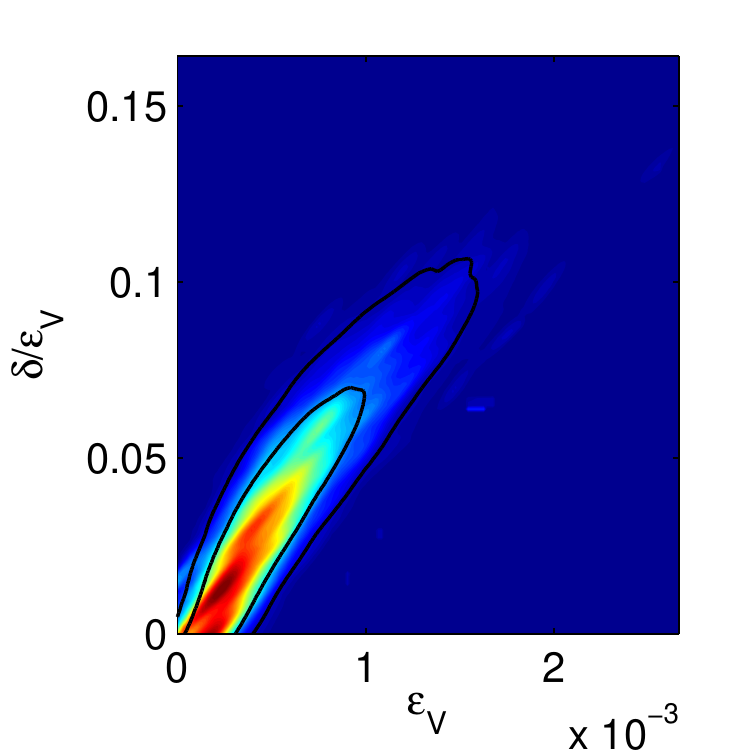}}
\subfigure[$\;\;\sigma=0.9, k_0=0.002 \rm{Mpc}^{-1}$]{\label{sigma09-st-2}
\includegraphics[width=4cm]{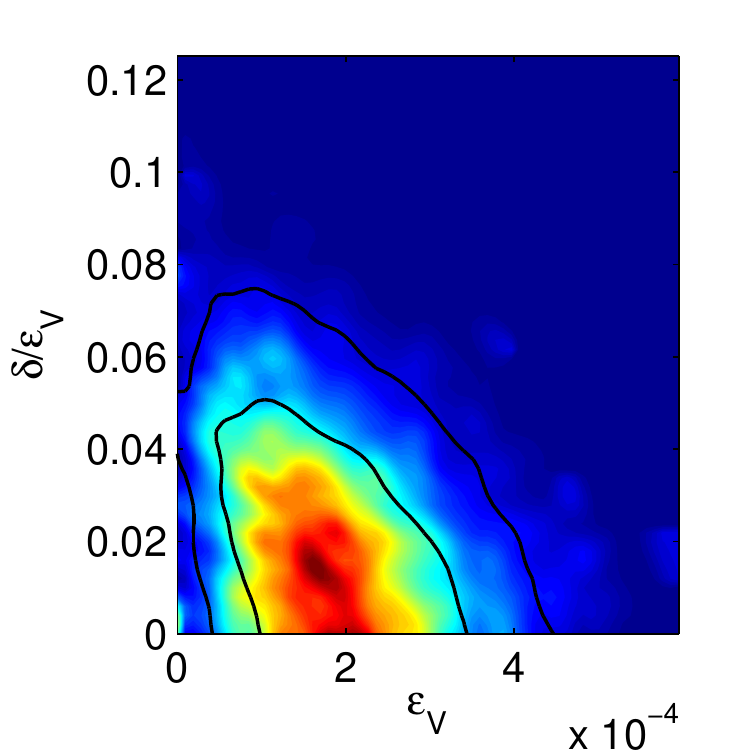}}
\subfigure[$\;\;\sigma=0.8, k_0=0.05\rm{Mpc}^{-1}$]{\label{sigma08-st-1}
\includegraphics[width=4cm]{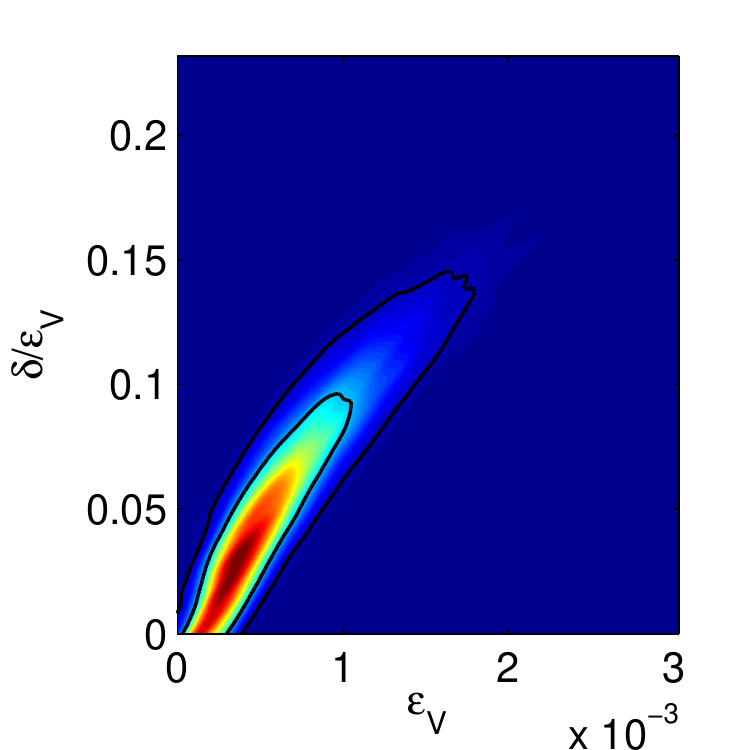}}
\subfigure[$\;\;\sigma=0.8, k_0=0.002 \rm{Mpc}^{-1}$]{\label{sigma08-st-2}
\includegraphics[width=4cm]{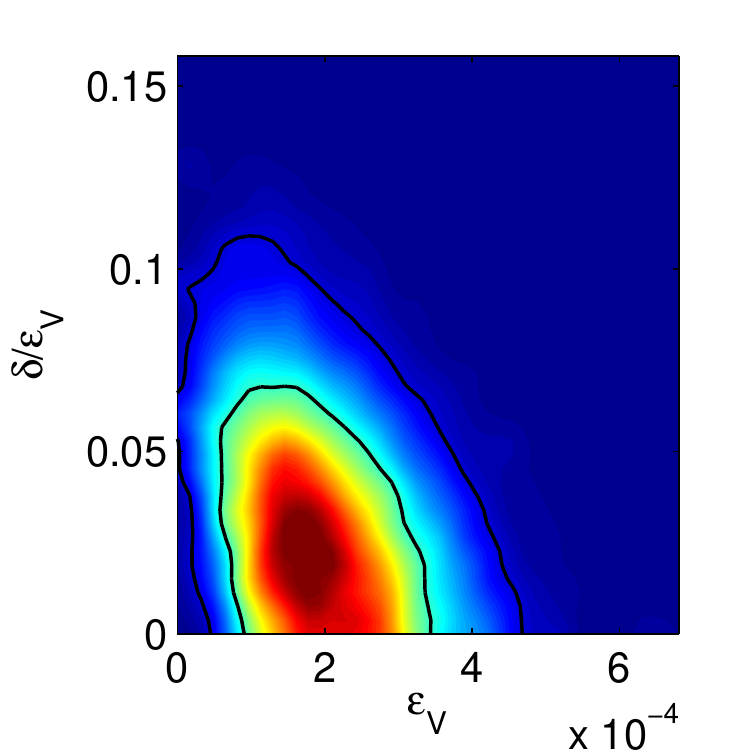}}
\\
\subfigure[$\;\;\sigma=0.6, k_0=0.05\rm{Mpc}^{-1}$]{\label{sigma06-st-1}
\includegraphics[width=4cm]{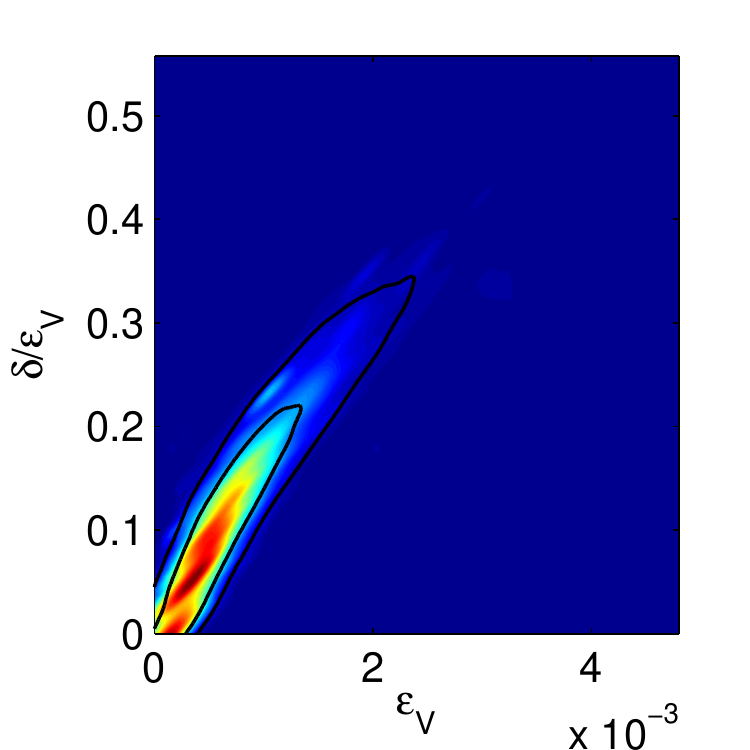}}
\subfigure[$\;\;\sigma=0.6, k_0=0.002 \rm{Mpc}^{-1}$]{\label{sigma06-st-2}
\includegraphics[width=4cm]{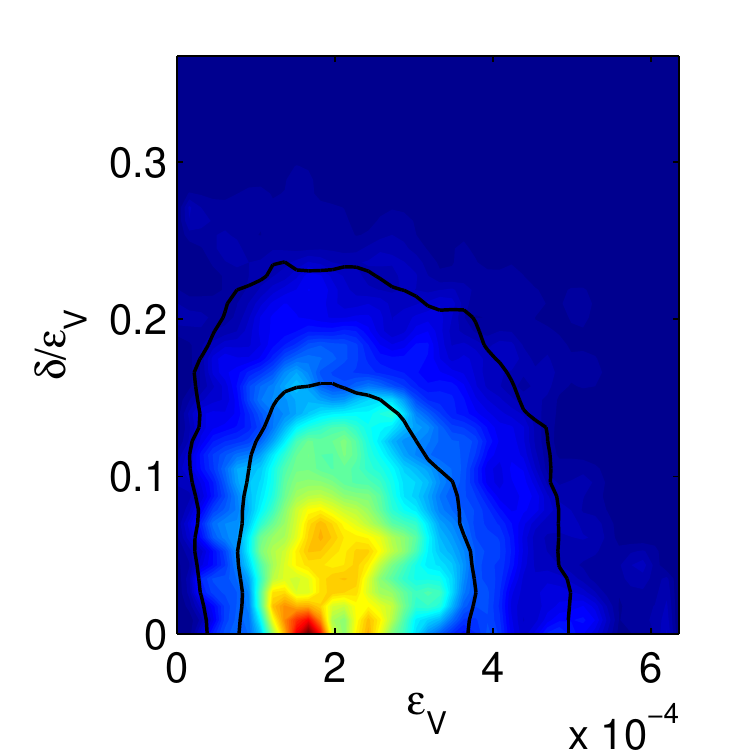}}
\subfigure[$\;\;\sigma=0.4, k_0=0.05\rm{Mpc}^{-1}$]{\label{sigma04-st-1}
\includegraphics[width=4cm]{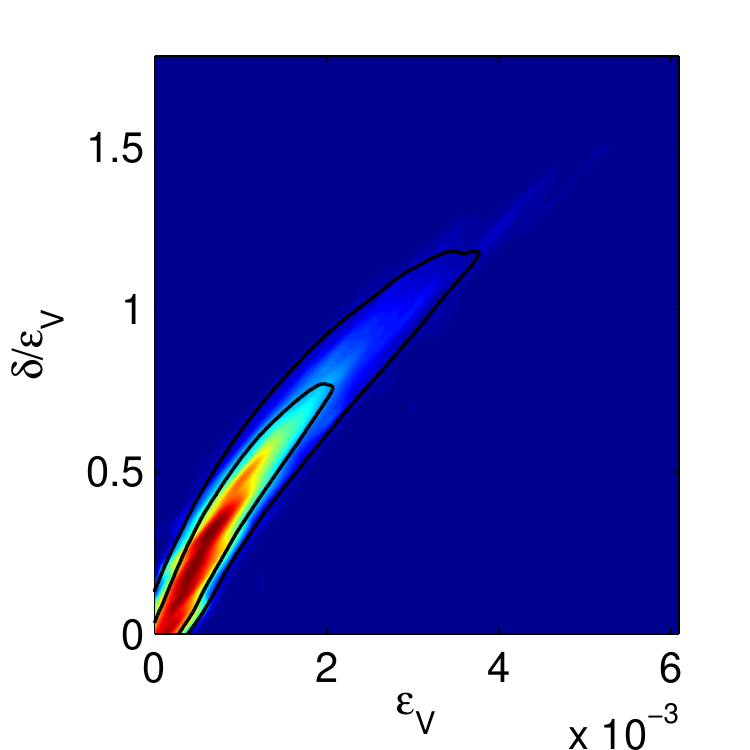}}
\subfigure[$\;\;\sigma=0.4, k_0=0.002 \rm{Mpc}^{-1}$]{\label{sigma04-st-2}
\includegraphics[width=4cm]{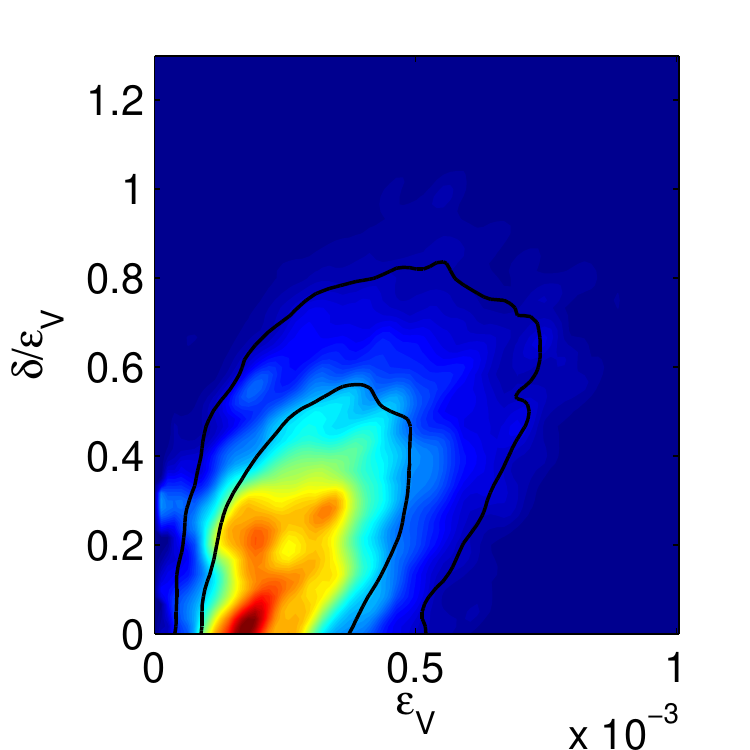}}
\caption{Two-dimensional marginalized distribution for the parameters $\delta/\epsilon_V$ and  $\epsilon_V$ for the inflation potential in Starobinsky inflation model at pivot scale $k_0=0.05\;\mbox{Mpc}^{-1} $ and $k_0=0.002\;\mbox{Mpc}^{-1}$, respectively. The internal and external lines correspond to the confidence levels of  $68\%$ and $95\%$, respectively.} \label{fig3}
\end{figure*}

\subsection{Detectability of inverse-volume corrections in future experiments}

Recengtly, various CMB missions have been proposed. These experiments will not only provide unprecedented precision of the CMB measurements, but also provide us a great chance to learn more about the inflation stage of the early universe. As the inflation happens at the energy scale not far from the Planck energy, one naturally expects that these cosmological observations could lead to a chance to learn something  about quantum gravity  in the early universe \cite{QGEs}. In this subsection, we are going to search for the potential observational signatures in the primordial inflationary spectra with the inverse-volume corrections.

In order to identify the observational effects of the inverse-volume corrections, a simple way is to find out whether we could distinguish these effects from some representative inflation models. In this paper we consider the inflation model with a power-law potential and Starobinsky potential as examples.

First, let us consider power-law potential. When the inverse-volume contributions vanish, we have $n_s=n_s(\epsilon_V)$ and $r=r(\epsilon_V)$. Thus it is easy to show that, up to the second-order of $\epsilon_V$, the relation
\cite{phi2}
\bqn
\lb{CNSTZ}
\Gamma_n (n_s, r) &\equiv& (n_s-1) + \frac{(2+n)r}{8n} \nb\\
&& + \frac{(3n^2 + 18n -4)(n_s-1)^2}{6(n+2)^2}   = 0, ~~~~
\eqn
holds precisely for a power-law potential. The results from Planck 2015    are $n_s = 0.968 \pm 0.006$ and $r_{0.002} < 0.11 (95 \%$ CL) \cite{Planck2015}, which yields $n_s \lesssim 1$.   In the forthcoming experiments, specially the   Stage
IV ones,  the errors  of the measurements on both $n_s$ and $r$ are  $ \sigma(n_s), \; \sigma (r) \le 10^{-3}$ \cite{S4-CMB}, which implies the error of the measurement of 
$\Gamma_n (n_s, r)$ is
\bq
\lb{Errors}
\sigma(\Gamma_n)   \le 10^{-3}.
\eq
Therefore, if any corrections to $n_s$ and $r$ lead to $\Gamma_n (n_s, r) \gtrsim 10^{-3}$, they should be within the range of detection of the current and forthcoming observations \cite{S4-CMB}. 

In particular,  when the inverse-volume corrections are taken into account ($\delta_{\text{Pl}} \not= 0$), we have $n_s = n_s(\epsilon_V, \epsilon_{\text{Pl}})$ and
$r = r(\epsilon_V,  \epsilon_{\text{Pl}})$, and  Eq.(\ref{CNSTZ})  is modified to,
\bq
\lb{CNSTZb}
\Gamma_n(n_s, r) = \mathcal{F}(\sigma) \frac{\delta(k)}{\epsilon_{V}},
\eq
where $\delta(k)\equiv \alpha_0 \epsilon_{\text{Pl}}H^{\sigma}$ and
\bqn
\lb{fs}
\mathcal{F}(\sigma) &=& -\frac{3^\sigma}{2^\sigma}\mathcal{K}_{-1}^{\star(s)}-\frac{\sigma ^2(\sigma -3)\alpha_0 }{18}(3D^\star_{\text{n}}\sigma-\sigma -3).\nb\\
\eqn
Clearly, the right-hand side of the above equation represents the quantum gravitational effects from the inverse-volume corrections. If it is  equal or greater than $ {\cal{O}}(10^{-3})$, these effects shall be within the detection of the current or  forthcoming experiments. It is interesting to note that the quantum gravitational effects are enhanced by a factor of $\epsilon_{V}^{-1}$,
which  is absent in \cite{Bojowald2011}.

In Fig.4, we show the curve of $\mathcal{F}(\sigma)$ vs $\sigma$, from which one finds that the absolute value of $\mathcal{F}(\sigma)$ drops down to very small values for $0.7<\sigma<0.8$ or for $\sigma$ is about zero.  
This will make the quantum gravitational effects very small, unless ${\delta(k)}/{\epsilon_{V}}$ is very large. Physically, this is not expected, and the above likelihood analysis also confirms it. Therefore, when $\sigma$ falls into these regions, the detectability  of the quantum gravitational effects from the inverse-volume corrections is rather low. However, except for these two particular regions, from the upper bounds listed in Table I and Table II, one finds that
\bqn
\mathcal{F}(\sigma) \frac{\delta}{\epsilon_V} \gtrsim \mathcal{O}(10^{-3}).
\eqn
Then, the quantum gravitational effects from the inverse-volume corrections could be well within the detectable range of the forthcoming experiments, specially the Stage IV ones.

For the Starobinsky potential, a relation similar to Eq.(\ref{CNSTZb}) for the power-law case also exists, but now takes the form, 
\bqn
\Gamma_S(n_s,r)\equiv (n_s-1)+\sqrt{\frac{r}{3}}=\mathcal{F}(\sigma) \frac{\delta}{\epsilon_V},
\eqn
where $\mathcal{F}(\sigma)$ is given by Eq.(\ref{fs}). Note that in this case $\Gamma_S(n_s,r) \propto \sqrt{r}$, in contrast to the power-law case in which we have $\Gamma_n(n_s,r) \propto {r}$. As a result, the sensibility of the measurement of $\Gamma_S(n_s,r)$ will be
\bq
\lb{GS}
\sigma\left(\Gamma_S\right) \simeq \sigma\left(\sqrt{r}\right) \simeq \mathcal{O}\left(10^{-1.5}\right) \sim \mathcal{O}\left(10^{-2}\right). 
\eq
Thus, in this case   the loop quantum effects is within the range of the detectability of the forthcoming generation of experiments, only when
\bqn
\left. \mathcal{F}(\sigma)\frac{\delta}{\epsilon_V}\right|_{Starobinsky} \geq \mathcal{O}\left(10^{-1.5}\right) \sim \mathcal{O}\left(10^{-2}\right).  ~~~~~~~~~~~~~~~~
\eqn
As can be seen from  Tables I, II and Fig.4, this is possible for some particular values of $\sigma$.

\begin{figure}
\lb{Fig4}
\includegraphics[width=8cm]{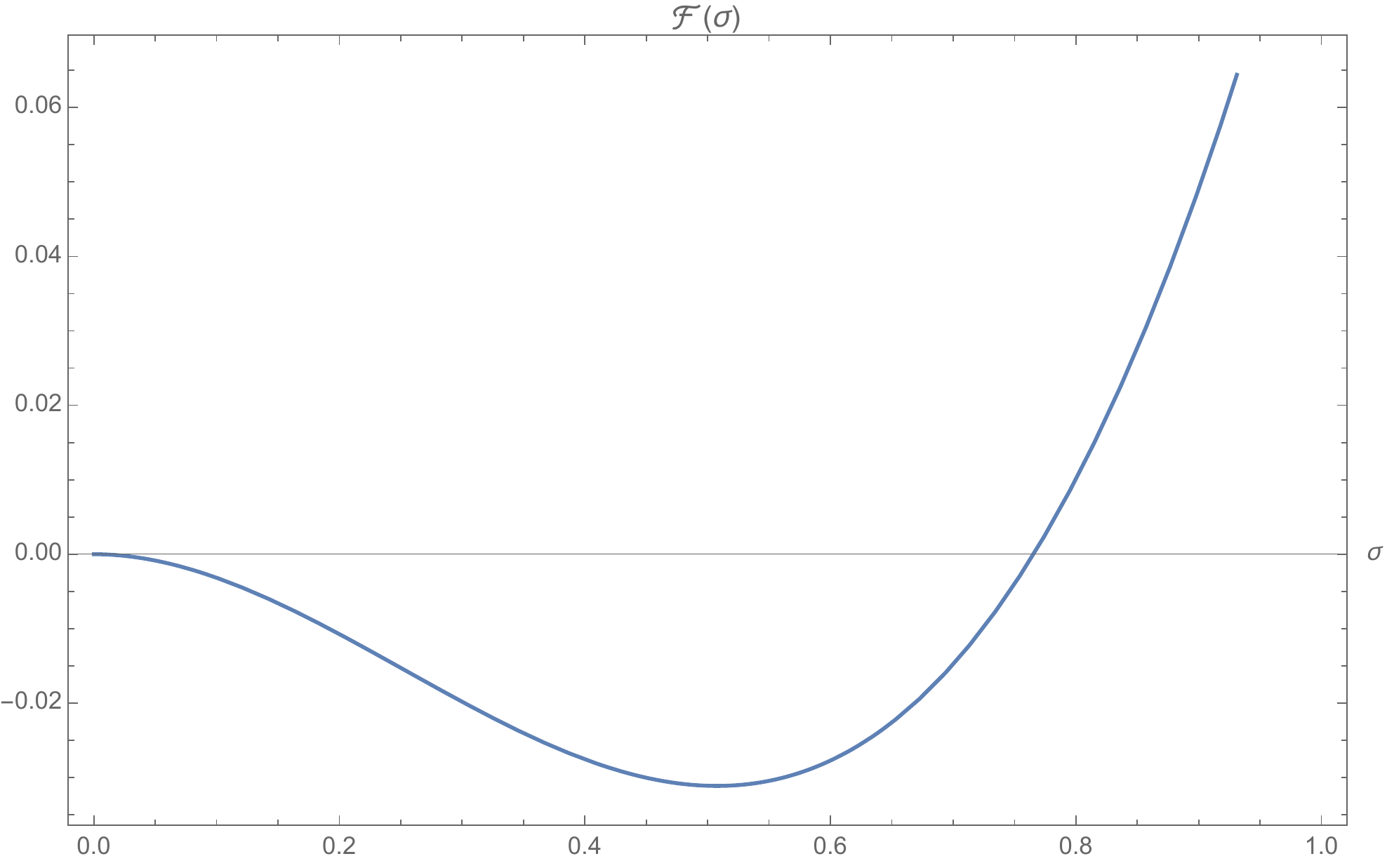}
\caption{The function $\mathcal{F}(\sigma)$ defined in Eq.(\ref{fs}) vs  $\sigma$.}
\end{figure}

\begin{table*}
\caption{Constarints on the parameters $\epsilon_V$, $\delta/\epsilon_V$, and $\delta(k_0)$ from Planck 2015 data at pivot scale $k_0=0.05 \mbox{Mpc}^{-1}$. }
\begin{ruledtabular}
\begin{tabular}{c|ccccc}
 & $V(\phi) $ &$ \epsilon_V $ (95\% C.L.)& $\epsilon_V$ (best fit) &$\frac{\delta}{\epsilon_V}$ (95\% C.L.)  & $\delta$ (95\% C.L.)\\
\hline
\multirow{5}{6em} {$\sigma=0.9$}  & $\phi$ & $\lesssim 0.0123$ & 0.00784  &   $\lesssim 0.0526$ & $\lesssim 4.1\times 10^{-4}$ \\
& $\phi^{2/3}$ &$\lesssim 0.00934$ &$ 0.00596$& $ \lesssim 0.0539$ & $\lesssim 3.2\times 10^{-4}$ \\
 & $\phi^{1/3}$ &$\lesssim 0.00528$ &$ 0.00337$&  $\lesssim 0.0499$ & $\lesssim  1.7\times 10^{-4}$ \\
 & $(1-e^{-\sqrt{2/3}\phi/M_{\text{Pl}}})^2$ &$\lesssim 0.00135$ &$ 0.00061 $&  $\lesssim 0.0901$ & $\lesssim  5.5\times 10^{-5}$ \\
 \hline
 \multirow{3}{6em} {$\sigma=0.8$}  & $\phi$ & $\lesssim 0.0126$ &$ 0.00803$ &   $\lesssim 0.0718$ & $\lesssim 5.8\times 10^{-4} $ \\
& $\phi^{2/3}$ &$\lesssim 0.00936$ &$ 0.00595 $ &   $\lesssim 0.0687$ & $\lesssim 4.1\times 10^{-4} $  \\
 & $\phi^{1/3}$ &$\lesssim 0.00553$ &$ 0.00345 $ &   $\lesssim 0.0702$ & $\lesssim 2.4\times 10^{-4} $\\
 & $(1-e^{-\sqrt{2/3}\phi/M_{\text{Pl}}})^2$ &$\lesssim 0.00149$ &$ 0.00067$&  $\lesssim 0.126$ & $\lesssim  8.4 \times 10^{-5}$ \\
 \hline
 \multirow{3}{6em} {$\sigma=0.7$}  & $\phi$ & $\lesssim 0.0133$ &$ 0.00817$ &   $\lesssim 0.106$ & $\lesssim  8.7\times 10^{-4} $  \\
& $\phi^{2/3}$ &$\lesssim 0.0967$ &$ 0.00608$ &   $\lesssim 0.0977$ & $\lesssim 5.9\times 10^{-4} $  \\
 & $\phi^{1/3}$ &$\lesssim 0.00563$ &$ 0.00354 $ &   $\lesssim 0.0959$ & $\lesssim 3.4\times 10^{-4} $ \\
  & $(1-e^{-\sqrt{2/3}\phi/M_{\text{Pl}}})^2$ &$\lesssim 0.00161$ &$ 0.00070$&  $\lesssim 0.180$ & $\lesssim  1.3 \times 10^{-4}$ \\
 \hline
 \multirow{3}{6em} {$\sigma=0.6$}  & $\phi$ & $\lesssim 0.0132$ &$ 0.00821$ &   $\lesssim 0.145$ & $\lesssim 1.2\times 10^{-3} $  \\
& $\phi^{2/3}$ &$\lesssim 0.0992$ &$ 0.00624$ &   $\lesssim 0.145$ & $\lesssim 9.0\times 10^{-4} $  \\
 & $\phi^{1/3}$ &$\lesssim 0.00578$ &$ 0.00361$ &   $\lesssim 0.142$ & $\lesssim 5.1\times 10^{-4} $  \\
  & $(1-e^{-\sqrt{2/3}\phi/M_{\text{Pl}}})^2$ &$\lesssim 0.00191$ &$ 0.00081$&  $\lesssim 0.292$ & $\lesssim  2.4 \times 10^{-4}$  \\
 \hline
  \multirow{3}{6em} {$\sigma=0.5$}  & $\phi$ & $\lesssim 0.00815$ &$0.00515 $ &   $\lesssim 0.154$ & $\lesssim 8.0\times 10^{-4} $ \\
& $\phi^{2/3}$ &$\lesssim 0.00621$ &$0.00433$ &   $\lesssim 0.162$ & $\lesssim 7.0\times 10^{-4} $ \\
 & $\phi^{1/3}$ &$\lesssim 0.00362$ &$ 0.00250$ &   $\lesssim 0.163$ & $\lesssim 4.1\times 10^{-4} $ \\
   & $(1-e^{-\sqrt{2/3}\phi/M_{\text{Pl}}})^2$ &$\lesssim 0.00369$ &$ 0.00021$&  $\lesssim 0.151$ & $\lesssim  2.4 \times 10^{-4}$  \\
 \hline
   \multirow{3}{6em} {$\sigma=0.4$}  & $\phi$ & $\lesssim 0.0148$ &$ 0.00893 $ &   $\lesssim 0.439$ & $\lesssim 3.9\times 10^{-3}$  \\
& $\phi^{2/3}$ &$\lesssim 0.0108$ &$ 0.00666 $ &   $\lesssim 0.415$ & $\lesssim 2.8\times 10^{-3} $  \\
 & $\phi^{1/3}$ &$\lesssim 0.00617$ &$ 0.00378 $ &   $\lesssim 0.386$ & $\lesssim 1.5\times 10^{-3}$ \\
 & $(1-e^{-\sqrt{2/3}\phi/M_{\text{Pl}}})^2$ &$\lesssim 0.0029$ &$ 0.00120$&  $\lesssim 1.03$ & $\lesssim  1.2 \times 10^{-4}$ \\
\end{tabular}
\end{ruledtabular}
\end{table*}

\begin{table*}
\caption{Constarints on the parameters $\epsilon_V$, $\delta/\epsilon_V$, and $\delta(k_0)$ from Planck 2015 data at pivot scale $k_0=0.002 \mbox{Mpc}^{-1}$.}
\begin{ruledtabular}
\begin{tabular}{c|cccccc}
 & $V(\phi) $ &$ \epsilon_V $ (95\% C.L.)& $\epsilon_V$ (best fit) &$\frac{\delta}{\epsilon_V}$ (95\% C.L.)  & $\delta$ (95\% C.L.) \\
\hline
\multirow{5}{6em} {$\sigma=0.9$}  & $\phi$ & $\lesssim 0.00680$ & 0.00417  &   $\lesssim 0.0663$ & $\lesssim 2.8\times 10^{-4}$\\
& $\phi^{2/3}$ &$\lesssim 0.00509$ &$ 0.00313 $& $ \lesssim 0.0648$ & $\lesssim 2.0 \times 10^{-4}$ \\
 & $\phi^{1/3}$ &$\lesssim 0.00284$ &$ 0.00177$&  $\lesssim 0.0654$ & $\lesssim  1.2\times 10^{-4}$\\
 & $(1-e^{-\sqrt{2/3}\phi/M_{\text{Pl}}})^2$ &$\lesssim 0.00036$ &$ 0.00018$&  $\lesssim 0.0601$ & $\lesssim  1.1\times 10^{-5}$  \\
 \hline
 \multirow{5}{6em} {$\sigma=0.8$}  & $\phi$ & $\lesssim 0.00680$ &$ 0.00430 $ &   $\lesssim 0.0866$ & $\lesssim 3.7\times 10^{-4} $  \\
& $\phi^{2/3}$ &$\lesssim 0.00514$ &$ 0.00320 $ &   $\lesssim 0.0884$ & $\lesssim 2.8 \times 10^{-4} $  \\
 & $\phi^{1/3}$ &$\lesssim 0.00292$ &$ 0.00184 $ &   $\lesssim 0.0878$ & $\lesssim 1.6 \times 10^{-4} $\\
 & $(1-e^{-\sqrt{2/3}\phi/M_{\text{Pl}}})^2$ &$\lesssim 0.00038$ &$ 0.00019$&  $\lesssim 0.0842 $ & $\lesssim  1.6 \times 10^{-5}$ \\
 \hline
 \multirow{5}{6em} {$\sigma=0.7$}  & $\phi$ & $\lesssim 0.00684$ &$ 0.00452$ &   $\lesssim 0.127$ & $\lesssim  5.7 \times 10^{-4} $ \\
& $\phi^{2/3}$ &$\lesssim 0.00522$ &$ 0.00338$ &   $\lesssim 0.127$ & $\lesssim 4.3 \times 10^{-4} $ \\
 & $\phi^{1/3}$ &$\lesssim 0.00299$ &$ 0.00193 $ &   $\lesssim 0.125$ & $\lesssim 2.4 \times 10^{-4} $  \\
  & $(1-e^{-\sqrt{2/3}\phi/M_{\text{Pl}}})^2$ &$\lesssim 0.00039$ &$ 0.00020$&  $\lesssim 0.121 $ & $\lesssim  2.4 \times 10^{-5}$ \\
 \hline
 \multirow{5}{6em} {$\sigma=0.6$}  & $\phi$ & $\lesssim 0.00725$ &$ 0.00482 $ &   $\lesssim 0.191$ & $\lesssim 9.2\times 10^{-4} $  \\
& $\phi^{2/3}$ &$\lesssim 0.00526$ &$ 0.00358 $ &   $\lesssim 0.193 $ & $\lesssim 6.9 \times 10^{-4} $  \\
 & $\phi^{1/3}$ &$\lesssim 0.00300$ &$ 0.00204 $ &   $\lesssim 0.191$ & $\lesssim 3.9\times 10^{-4} $  \\
  & $(1-e^{-\sqrt{2/3}\phi/M_{\text{Pl}}})^2$ &$\lesssim 0.00044 $ &$ 0.00023 $&  $\lesssim 0.195 $ & $\lesssim 4.5 \times 10^{-5}$ \\
 \hline
  \multirow{5}{6em} {$\sigma=0.5$}  & $\phi$ & $\lesssim 0.00696$ &$0.00435 $ &   $\lesssim 0.492$ & $\lesssim 2.1\times 10^{-3} $ \\
& $\phi^{2/3}$ &$\lesssim 0.00514$ &$0.00320$ &   $\lesssim 0.509$ & $\lesssim 1.6\times 10^{-3} $ \\
 & $\phi^{1/3}$ &$\lesssim 0.00295$ &$ 0.00186$ &   $\lesssim 0.486$ & $\lesssim 9.0\times 10^{-4} $ \\
   & $1-e^{-\sqrt{2/3}\phi/M_{\text{Pl}}}$ &$\lesssim 0.00028$ &$ 0.00012$&  $\lesssim 0.247$ & $\lesssim  3.0\times 10^{-5}$ \\
 \hline
   \multirow{3}{6em} {$\sigma=0.4$}  & $\phi$ & $\lesssim 0.00808$ &$ 0.00566 $ &   $\lesssim 0.583$ & $\lesssim 3.3\times 10^{-3}$ \\
& $\phi^{2/3}$ &$\lesssim 0.00599$ &$ 0.00420 $ &   $\lesssim 0.576$ & $\lesssim 2.4 \times 10^{-3} $  \\
 & $\phi^{1/3}$ &$\lesssim 0.00341 $ &$ 0.00241 $ &   $\lesssim 0.570$ & $\lesssim 1.4 \times 10^{-3}$ \\
 & $(1-e^{-\sqrt{2/3}\phi/M_{\text{Pl}}})^2$ &$\lesssim 0.00063$ &$ 0.00033$&  $\lesssim 0.688$ & $\lesssim  2.3 \times 10^{-4}$ \\
\end{tabular}
\end{ruledtabular}
\end{table*}

\section{Conclusions}

The uniform asymptotic approximation method provides a powerful, systematically improvable, and error control approach to construct accurate analytical solutions of inflationary perturbations. It has been proved to be a very effective method by applying it to inflation models with nonlinear dispersion relations \cite{Zhu1, Zhu2, Uniform3}, $k$-inflation \cite{Uniform4}, and  holonomy and inverse-volume corrections from LQC \cite{Zhu3}. In this paper, we provide a new  approach to calculate the inflationary spectra by using the uniform asymptotic approximation method, after the inverse-volume corrections from LQC are taken into account. This new approach allows us to consider the case with any given value of $\sigma$, a free parameter appearing in the inverse-volume corrections \cite{Bojowald2011}. 

Previously, a different approach was taken, and it allowed us only to calculate the power spectra in the case where $\sigma$ is an integer \cite{Zhu3}. In this sense, the current work is a natural generalization of our previous work \cite{Zhu3}, so that the power spectra, spectral indices,  runnings, and tensor-to-scalar ratio of the power spectra are explicitly calculated up to the second-order of the slow-roll parameters and third-order of the parameter $\lambda^{-1}$ introduced in the method for the inverse-volume corrections from LQC for any given $\sigma$. Up to the third-order of $\lambda^{-1}$, the upper error bounds  are less than $0.15\%$ \cite{Uniform3}. In addition, as a
consistent check, when $\sigma$ is an integer, both approaches give the same results, as it is expected. Considering the fact that very heavy mathematical calculations are highly involved in both of the approaches, it is still a bit of surprising that they give precisely the same expressions.  This is  thanks to a careful analysis of the problem presented in Appendix B.

We also apply  the COSMOMC code developed by us previously to carry out the CMB likelihood analysis, in order to search for the observational constraints on the inverse-volume quantum corrections from the latest release Planck 2015 data.  Via such analysis we place observational bounds on both the inverse-volume correction parameter $\delta$ and the slow-roll parameter $\epsilon_V$  for the power-law potential as well as for  the Starobinsky potential. The constraints depend on the values of  $\sigma$ and the values of the pivot scale $k_0$. In particular, when $\sigma$ is larger than one, the constraints become so tight that quantum gravitational effects from the inverse-volume corrections are very small, and cannot be detected in the current and forthcoming experiments.  This is consistent with the previous conclusions obtained in \cite{Zhu3, Bojowald2011}. In addition, using these constraints we also show that the inverse-volume quantum corrections in primordial spectra might be potentially detectable or can be tightly constrained by the forthcoming  experiments, such as the Stage IV ones \cite{S4-CMB}, provided that
$\sigma \leq 1$.

{ Finally, we note that, when  both inverse-volume and holonomy corrections are  taken into account simultaneously, anomaly-free perturbations  were studied in \cite{CLB14}, and found that it is still possible to close the algebra of the constraints. In particular, it was shown that even in the case   studied in this paper  \cite{Bojowald01,Bojowald2011},   
the holonomy corrections will in general be affected by the inverse-volume ones \cite{CLB14}, although  in the stage of slow-roll inflation such effects are expected not to be large \cite{Zhu3}.  
In addition, a generic feature of loop quantum cosmology is the replacement of the big bang singularity by a non-singular bounce \cite{AB}.  As noted previously, the results presented here are valid only in the later slow-roll inflationary phase. It would be extremely interesting and important to understand the effects of such a pre-slow-roll phase.  Note that our {\em uniform asymptotic approximation method} is valid not only for slow-roll inflationary models, but also for non-slow-roll ones. So, in principle one can apply our general formulas of perturbations directly to such studies. We hope to come back to this important issue soon in another occasion.}

\section*{Acknowledgements}

 T.Z. would like to express his gratitude to Dr. Yungui Gong for the valuable discussions. Some of the preliminary runnings of COSMOMC code were setup and carried out when T.Z. was visiting the Kavli Institute for Theoretical Physics,  the Chinese Academy of Sciences in Beijing, China.  This work is supported in part by Ci\^encia Sem Fronteiras, No. 004/2013 - DRI/CAPES, Brazil (A.W.); Chinese NSFC No. 11375153 (A.W.), No. 11173021 (A.W.), 
 No. 11205133 (Q.W.),  No. 11047008 (T.Z.), No. 11105120 (T.Z.), and No. 11205133 (T.Z.).

\section*{Appendix A: Convergence of the error control functions}
\renewcommand{\theequation}{A.\arabic{equation}} \setcounter{equation}{0}

\subsection*{Liouville-Green approximation and convergence of the error control function near poles}

In this section, let us discuss the case with a general $\sigma$ with the restriction $0<\sigma \leq 6$. Let us consider  the equation, 
\bqn\lb{eom-sigma}
\frac{d^2\mu_k(y)}{dy^2}=\left\{\lambda^2 \hat g(y)+q(y)\right\}\mu_k(y),
\eqn
where $y=-k\eta$, and
\bqn\lb{gq}
\lambda^2 \hat g(y)+q(y)=\frac{\nu^2-1/4}{y^2}-1-\chi \epsilon_{\text{Pl}} \kappa y^\sigma+m \epsilon_{\text{Pl}} \kappa y^{\sigma-2}.\nb\\
\eqn
Obviously, the above equations have  
 two poles, one is at $y=0^{+}$ and another is at $y=+\infty$. Near the two poles, the approximate solutions of the above equation can be constructed by the Liouville-Green (LG) approximation. 

Specifically, {\em near the pole $y=0^{+}$, suppose $\lambda^2 \hat g(y)$ being a real and twice continuously differentiable function, $q(x)$ a continuous real function}, then the LG solution takes the form
\bqn
\mu_k(y)&=&\frac{c_{+}}{g(y)^{1/4}} e^{\int^y \sqrt{\lambda^2 \hat g(y)} dy} (1+\epsilon_1^{+})\nb\\
&&+\frac{d_{+}}{g(y)^{-1/4}} e^{-\int^y \sqrt{\lambda^2 \hat g(y)}dy} (1+\epsilon_{2}^{+}),
\eqn
where $\epsilon_1^{+}$ and $\epsilon_2^{+}$ represent the errors of the approximate solution, which is characterized by the error control function $\mathcal{F}(y)$ near the pole $0^{+}$. To analyze the behavior of $\mathcal{F}(y)$, let us first write down its expression 
\bqn
\mathcal{F}(\xi) =\int \left\{\frac{1}{g^{1/4}} \frac{d^2}{dy^2}\left(\frac{1}{g^{1/4}}\right)-\frac{q}{g^{1/2}}\right\}dy.
\eqn
The LG approximation is meaningful only when the associated error control function $\mathcal{F}(y)$ is convergent near the poles. This condition also provides a guidance for how to determine the splitting of  $\lambda^2 \hat g(y)+q(y)$. According to the result \cite{Olver1974} (Chapter 6, Sec 4.1, page 200), the error control function $\mathcal{F}(\xi)$ is convergent if
\bqn\lb{irregular singularity}
\lambda^2 \hat g(y) \sim \frac{c_0}{y^{2 d_0+2}},\;\;\;\;q(y) \sim \mathcal{O}\left\{\f{1}{y^{d_0-e_0+2}}\right\},
\eqn
when $y\to 0^{+}$, where $c_0$, $d_0$, and $e_0$ are positive constants. Note that we also require that the first relation is twice differentiable. Here we emphasize that the above condition (\ref{irregular singularity}) includes the case when the equation (\ref{eom-sigma}) has an {\em irregular singularity} at $y=0^+$ of rank $d_0$. For a regular singularity, the function $\lambda^2 \hat g(y)$ and $q(y)$ should be expanded in a series of the form
\bqn
\lambda^2 \hat g(y)=y^{-2} \sum_{s=0}^{+\infty} g_s y^s,\;\;\;q(y)=y^{-2} \sum_{s=0}^{+\infty} q_s y^s,
\eqn
then it is easy to show that the error control function is convergent only when
\bqn
q_0=-\frac{1}{4}.
\eqn

However, in the case with $\lambda^2 \hat g(y)+q(y)$ given by Eq.(\ref{gq}), the pole $y=0^+$ is neither an irregular nor a regular singularity. In this case, in order to study the error control function, let us assume that the function $\lambda^2 \hat g(y)$ and $q(y)$ can be expanded near the pole $y=0^{+}$ in the form
\bqn\lb{new_expansion}
\lambda^2 \hat g(y) =y^{-i} \sum_{s=0}^{\infty} (g_s+g_{\sigma s}y^{\sigma}) y^s,\\
q(y) = y^{-j} \sum_{s=0}^{\infty} (q_s+q_{\sigma s}y^{\sigma})y^{s}.
\eqn
Thus it is easy to show that
\bqn
&&g_{i-2}+q_{j-2}=\nu^2-\frac{1}{4},\nb\\
&&g_{i}+q_{j}=-1,\nb\\
&&g_{s}+q_{s+j-i}=0,\;\;\;\;\text{if}\;\;s\neq i-2,\; i,
\eqn
and
\bqn
&&g_{\sigma (i-2)}+q_{\sigma (j-2)}=m \epsilon_{\text{Pl}} \kappa,\nb\\
&&g_{\sigma i}+q_{\sigma j}=-\chi \epsilon_{\text{Pl}} \kappa,\nb\\
&&g_{\sigma s}+q_{\sigma (s+j-i)}=0, \;\;\;\;\text{if}\;\;s\neq i-2,\; i.
\eqn
Putting these expansions in the integrand of the error control function, we find
\bqn
&&\frac{1}{g^{1/4}} \frac{d^2}{dy^2}\left(\frac{1}{g^{1/4}}\right)-\frac{q}{g^{1/2}}\nb\\
&&~~~~=\frac{5}{16} \frac{g'^2}{g^{5/2}}-\frac{1}{4} \frac{g''}{g^{3/2}}-\frac{q}{g^{1/2}},
\eqn
in which we have
\bqn
\frac{5}{16} \frac{g'^2}{g^{5/2}} &\simeq& \frac{5}{16} y^{\frac{i}{2}-2} \left\{\sum_{s=0}^{\infty} [(s-i)g_s+(s+\sigma-p)g_{\sigma s}y^{\sigma}]y^s\right\}^2\nb\\
&&~~~~\times \left\{\sum_{s=0}^{\infty} (g_s+g_{\sigma s }y^\sigma)y^s\right\}^{-5/2}\nb\\
&&\simeq \frac{5}{16} y^{-2+i/2} \left[4 g_0^{-1/2}+\mathcal{O}(y^{\sigma})\right],\nb\\
\eqn
similarly,
\bqn
-\frac{1}{4} \frac{g''}{g^{3/2}} &\simeq &-\frac{1}{4} y^{-2+i/2} \left[6g_0^{-1/2}+\mathcal{O}(y^{\sigma})\right],\nb\\
\eqn
and
\bqn
-\frac{q}{g^{1/2}} &\simeq& -y^{-j+i/2} \left[q_0 g_0^{-1/2}+\mathcal{O}(y^{\sigma})\right].
\eqn
In the following, let us discuss the error control function case by case. 
\begin{itemize}
\item $i>2$. In this case, the error control function is convergent only if $j<\frac{i}{2}+1$, i.e.,
\bqn
i>2,\;\;j<\frac{i}{2}+1.
\eqn
\item $i=2$. In this case, one requires $j=2$ and the error control function reads
\bqn
&&\frac{5}{16} \frac{g'^2}{g^{5/2}}-\frac{1}{4} \frac{g''}{g^{3/2}}-\frac{q}{g^{1/2}}\nb\\
&& \sim \left(\frac{5}{16}\times 4g_0^{-1/2}-\frac{1}{4}\times 6 g_0^{-1/2}-q_0 g_0^{-1/2}\right)y^{-1}\nb\\
&&= -g_0^{-1/2}\left(q_0+\frac{1}{4}\right)y^{-1}+\mathcal{O}(y^{\sigma-1}),
\eqn
which gives
\bqn
i=2,\;j=2,\;q_0=-\frac{1}{4}.
\eqn
\item $i<2$. In this case, the error control function cannot be convergent.
\end{itemize}
In review of  the above analysis, we find that the error control function $\mathscr{F}(\xi)$ is convergent if we choose
\bqn\lb{gqcondition}
\lambda^2 \hat g(y)&=&\frac{\nu^2}{y^2}-1-\chi \epsilon_{\text{Pl}}\kappa y^{\sigma}+m\epsilon_{\text{Pl}} \kappa y^{\sigma-2},\\
q(y)&=&- \frac{1}{4y^2}.
\eqn

Similarly, near the pole $y=+\infty$, the LG approximate solution takes the form
\bqn
\mu_k(y)&=&\frac{c_{-}}{g(y)^{1/4}} e^{i\int^y \sqrt{\lambda^2 \hat g(y')}dy'}(1+\epsilon_{1}^{-})\nb\\
&&+\frac{d_{-}}{g(y)^{1/4}}e^{-i\int^y \sqrt{\lambda^2 \hat g(y')}dy'}(1+\epsilon_2^{-}),
\eqn
where $\epsilon_{1}^{-}$ and $\epsilon_2^{-}$ represent the errors of the approximate solutions, which are characterized by the error control function $\mathcal{F}(y)$ near the pole $+\infty$. Again, according to the result given in \cite{Olver1974} (Chapter 6, Sec 4.2, page 201), the error control function is convergent if
\bqn\lb{condition_infty}
\lambda^2 \hat g(y) \sim c_0 y^{2d_0-2},\;\;\;q(y)\sim \mathcal{O}(y^{d_0-e_0-2}),
\eqn
where $c_0$, $d_0$, and $e_0$ are another set of positive constants. It is easy to see that the condition of Eq.(\ref{condition_infty}) is  included in the conditions of $\lambda^2 \hat g(y)$ and $q(y)$ near $y=0^+$. Thus the associated error control function is convergent near the pole $y=+\infty$, if $\lambda^2 \hat g(y)$ and $q(y)$ are given by Eq.(\ref{gqcondition}).

\subsection*{Convergence of $\mathscr{H}(+\infty)$ near turning point}

The error control function $\mathscr{H}(\xi)$ given by Eq.(\ref{H}) is expected to be convergent near the turning point. However, A quick look at Eq.(\ref{H}) seemingly tells that there are apparent divergencies in the error control function when evaluating it in the limit $y\to \bar y_0$. In this appendix, we will show that these apparent singularities cancel each other, and finite results indeed exist. Near the turning point $y=\bar y_0$, one usually has
\bqn
q(\bar y_0) &\neq& 0,\nb\\
\lambda^2 \hat g(y) &\simeq& -g'(\bar y_0)(\bar y_0-y)+\frac{1}{2}g''(\bar y_0)(\bar y_0-y)^2.\nb
\eqn
Then the first part of the error control function can be expanded as
\bqn\lb{first_part}
&&\frac{5}{36}\left\{\int_{\bar y_0}^{0}\sqrt{g(y')}dy'\right\}^{-1}-\frac{5}{36}\left\{\int_{\bar y_0}^{\bar y_0-\varepsilon}\sqrt{g(y')}dy'\right\}^{-1}\nb\\
&&~~\simeq -\frac{5}{24} \frac{1}{\sqrt{-g'(\bar y_0)}} \frac{1}{\varepsilon^{3/2}}+\frac{1}{32}\frac{g''(\bar y_0)}{(-g'(\bar y_0))^{3/2}}\frac{1}{\sqrt{\varepsilon}},\nb\\
\eqn
where $\varepsilon$ is a positive and small quantity, which is introduced to represent the divergences in the expressions. Similarly, for the second part it is also easy to find
\bqn\lb{second_part}
&&-\int_{\bar y_0-\varepsilon}^y \left\{\frac{q}{g}-\frac{5g'^2}{16g^3}+\frac{g''}{4g^2}\right\}\sqrt{g}dy'\nb\\
&& ~~~\simeq\frac{5}{24} \frac{1}{\sqrt{-g'(\bar y_0)}} \frac{1}{\varepsilon^{3/2}}-\frac{1}{32}\frac{g''(\bar y_0)}{(-g'(\bar y_0))^{3/2}}\frac{1}{\sqrt{\varepsilon}}\nb\\
&&~~~~~+\mathcal{O}(\sqrt{\varepsilon}).
\eqn
Obviously, combining Eq.(\ref{first_part}) and Eq.(\ref{second_part}), it is easy to show that the singularities exactly cancell each other, and the convergence of the error control function $\mathscr{H}(+\infty)$ is fulfilled.

\section*{Appendix B: Expansions of Integrals with a small parameter}
\renewcommand{\theequation}{B.\arabic{equation}} \setcounter{equation}{0}

\subsection*{B.1 An example for expansion of an integral}

Let us first consider a simple integral in the form
\bqn\lb{example}
I[\epsilon]=\int_{0}^{1-\epsilon} (1-\epsilon+x)^{\alpha} dx,
\eqn
where $\epsilon$ is a small constant, $\alpha$ is a constant, and one requires $\alpha>-1$ to ensure the above integral to be convergent. The above integral can be performed exactly, which yields
\bqn
I[\epsilon]=\frac{(1-\epsilon)^{1+\alpha}}{1+\alpha}.
\eqn
Then consider the Taylor series of the above result about $\epsilon=0$. We find
\bqn\lb{Taylor_app}
I[\epsilon] \simeq \frac{1}{1+\alpha}-\epsilon+\frac{\alpha}{2} \epsilon^2+\mathcal{O}(\epsilon^3).
\eqn

However, in practice the integral we would like to perform may not be as simple as that given in Eq.(\ref{example}). In this case, one approach to carry out the calculation is to do the Taylor expansion of the integrand first, instead of doing the integral directly. For the integral of Eq.(\ref{example}), a direct Taylor expansion about $\epsilon=0$ gives
\bqn
I[\epsilon]\simeq \int_{0}^{1} (1-x)^{\alpha}dx-\epsilon \int_0^1 \alpha (1-x)^{\alpha-1}dx\nb\\
+\frac{\epsilon^2}{2} \int_0^1 \alpha(\alpha-1)(1-x)^{\alpha-2}dx.
\eqn
A problem arises immediately: the second and third integrals in the above expressions are not convergent if $\alpha<0$. 
Obviously,  this is not consistent with Eq.(\ref{Taylor_app}), which is valid for all $\alpha>-1$. Inspecting the above expression, one notices that the discrepancy comes from the fact that the Taylor series of the integrand breaks down when $x \to 1$. We can see this clearly if we consider the following expansions,
\bqn
\sqrt{1-x-\epsilon}\simeq \sqrt{1-x}-\frac{\epsilon}{2\sqrt{1-x}}-\frac{\epsilon^2}{8(1-x)^{3/2}}+\mathcal{O}(\epsilon^3),\nb\\
\eqn
which obviously breaks down when $x\to 1$. Thus, we conclude that the approximation of the integral in Eq.(\ref{Taylor_app}) depends on the value of parameter $\alpha$ and also the order of the expansion. For $\alpha>1$, we can expand the integral up to the order $\mathcal{O}(\epsilon^2)$, while for $\alpha>0$, we can only expand to the first order.

Now let us apply the above to the analysis of the expansions of $\int \sqrt{g(y')}dy'$ and the error control function $\mathscr{H}(+\infty)$. We first consider the integral of $\sqrt{g(y)}$, which can be expressed as
\bqn
\int_y^{\bar y_0} \frac{\sqrt{\bar y_0-y}}{y}\left(\cdots\cdots \right)dy'.
\eqn
Obviously, the above integral has the same structure as that of Eq.(\ref{example}) by taking $\alpha=\frac{1}{2}$. As a result one concludes that the Taylor expansion of the above integral  is only valid if we consider the approximation at the first order of $\epsilon_{\text{Pl}}$.  Similar analysis applies to the error control function $\mathscr{H}(+\infty)$. As shown in the last section, all the singularities arising in the integral when $y \to \bar y_0$ should be cancelled  each other. Thus, we can safely expand the error control function in terms of $\epsilon_{\text{Pl}}$ up to its first-order.

\subsection*{B.2 Expansions of $\int \sqrt{g(y')}dy'$ and the error control function $\mathscr{H}(\infty)$}

In order to expand an integral in terms of a small parameter $\epsilon$, let us consider the following formula
\bqn
I[a(\epsilon),b(\epsilon),\epsilon]=\int_{a(\epsilon)}^{b(\epsilon)} F[y',\epsilon]dy'.
\eqn
Now expanding the above integral in terms of $\epsilon$ yields
\bqn\lb{derint}
&&I[a(\epsilon),b(\epsilon),\epsilon]\nb\\
&&~~~\simeq \int_{a(0)}^{b(0)} F(y',0)dy'+\epsilon \int_{a(0)}^{b(0)} F_{,\epsilon}(y',0)dy'\nb\\
&&~~~~~+\epsilon \Big[b_{,\epsilon}(0)F(b_0,0)-a_{,\epsilon}(0)F(a_0,0)\Big].
\eqn
Using this formula, the integral (\ref{power_spectra}) can be expanded in terms of $\epsilon_{\text{Pl}}$ as
\bqn\lb{int_of_sqrt_g}
&&\int^{\bar y_0}_{y}\sqrt{\lambda^2 g(y')}dy'\nb\\
&&~~\simeq \int^{\bar \nu_0}_{y} \sqrt{g(y')}|_{\epsilon_{\text{Pl}}=0}dy'\nb\\
&&~~~~+\epsilon_{\text{Pl}}\int_{y}^{\bar \nu_0} \left.\frac{g_{,\epsilon_{\text{Pl}}}(y')}{2\sqrt{g(y')}}\right|_{\epsilon_{\text{Pl}}=0}dy'.
\eqn
Note that the upper limit of the integral, $\bar y_0 \simeq \bar \nu_0+\frac{1}{2}\bar \kappa_0 (\bar m_0 -\chi \bar \nu_0^{-2}) \bar \nu_0^{\sigma-1}  \epsilon_{\text{Pl}}$, is a function of $\epsilon_{\text{Pl}}$. 

Now let us turn to the error control function $\mathscr{H}(+\infty)$ in Eq.(\ref{H}), from which we can divide $\mathscr{H}(+\infty)$ into two parts. We first consider the first part, which is
\bqn
&&\frac{5}{36}\left.\frac{1}{I_1[y,\bar y_0,\epsilon_{\text{Pl}}}\right|_{\bar y_0-\varepsilon}^{0^+}\nb\\
&&~~~~~~~~~~~~=\frac{5}{36}\left\{\frac{1}{I_1[0,\bar y_0,\epsilon_{\text{Pl}}]}-\frac{1}{I_1[\bar y_0-\varepsilon,\bar y_0,\epsilon_{\text{Pl}}]}\right\},\nb\\
\eqn
where
\bqn\lb{B5}
I_1[0,\bar y_0,\epsilon_{\text{Pl}}]&=&\lim_{y\to 0^+}\int_{\bar y_0}^{y}\sqrt{g(y')}dy',\nb\\
I_1[\bar y_0-\varepsilon,\bar y_0,\epsilon_{\text{Pl}}]&=&\int_{\bar y_0}^{\bar y_0-\varepsilon }\sqrt{g(y')}dy',
\eqn
and similar to the last section, here $\varepsilon$ is a positive and small quantity, representing the divergences in the expressions . Using the formula (\ref{derint}), we find
\bqn
I_1[0,\bar y_0,\epsilon_{\text{Pl}}]&\simeq&\lim_{y\to 0^+} \int_{\bar \nu_0}^{y} \sqrt{g(y')}|_{\epsilon_{\text{Pl}}=0}dy'\nb\\
&&+\epsilon_{\text{Pl}}\Bigg\{\lim_{y\to0^+} \int_{\bar \nu_0}^{y} \left.\frac{g_{,\epsilon_{\text{Pl}}}(y')}{2\sqrt{g(y')}}\right|_{\epsilon_{\text{Pl}}=0}dy'\nb\\
&&~~~~~~-\bar y_{0,\epsilon_{\text{Pl}}} \sqrt{g(\bar \nu_0)}|_{\epsilon_{\text{Pl}}=0}\Bigg\}.\nb\\
\eqn
Note that we have $g(\bar \nu_0)|_{\epsilon_{\text{Pl}}=0}=0$. As we show in Eq.(\ref{sl_sqrt_g}), the above integral contains a divergent term $\ln \frac{y}{\bar \nu_0}$ in the limit $y\to 0^+$. Thus we have
\bqn\lb{I0}
\frac{5}{36}\frac{1}{I_1[0,\bar y_0,\epsilon_{\text{Pl}}]}=0.
\eqn
Similarly, for $I_1[\bar y_0-\varepsilon,\bar y_0,\epsilon_{\text{Pl}}]$, we have
\bqn
I_1[\bar y_0-\varepsilon,\bar y_0, \epsilon_{\text{Pl}}]&\simeq&\lim_{\varepsilon\to 0} \int_{\bar \nu_0}^{\bar \nu_0-\varepsilon}\sqrt{g(y')}|_{\epsilon_{\text{Pl}}=0}dy'\nb\\
&&+\epsilon_{\text{Pl}}\lim_{\varepsilon \to0^+} I_{1,\epsilon_{\text{Pl}}}[\bar \nu_0-\varepsilon,\bar \nu_0,0],\nb\\
\eqn
where
\bqn\lb{B9}
&&I_{1,\epsilon_{\text{Pl}}}[\bar \nu_0-\varepsilon,\bar \nu_0,0]\nb\\
&&~~~= \int_{\bar \nu_0}^{\bar \nu_0-\epsilon} \left.\frac{g_{,\epsilon_{\text{Pl}}}(y')}{2\sqrt{g(y')}}\right|_{\epsilon_{\text{Pl}}=0}dy'\nb\\
&&~~~~+\bar y_{0,\epsilon_{\text{Pl}}} \sqrt{g(\bar \nu_0-\varepsilon)}|_{\epsilon_{\text{Pl}}=0}-\bar y_{0,\epsilon_{\text{Pl}}} \sqrt{g(\bar \nu_0)}|_{\epsilon_{\text{Pl}}=0}.\nb\\
\eqn
Using this expression we find
\bqn
&&\frac{5}{36I_1[\bar y_0-\varepsilon,\bar y_0,\epsilon_{\text{Pl}}]}\nb\\
&&~~~\simeq \frac{5}{36I_1[\bar \nu_0-\varepsilon,\bar \nu_0,0]}-\epsilon_{\text{Pl}}\frac{5I_{1,\epsilon_{\text{Pl}}}[\bar \nu_0-\varepsilon,\bar\nu_0,0]}{36I_1[\bar \nu_0-\varepsilon,\bar \nu_0,0]^2 }.\nb\\
\eqn
Then combining with Eq.(\ref{I0}), the first part of the error control function can be calculated by using the following formula
\bqn
&&\frac{5}{36}\left.\frac{1}{I_1[y,\bar y_0,\epsilon_{\text{Pl}}}\right|_{y=\bar y_0-\varepsilon}^{y=0^+}\nb\\
&&~~~\simeq -\frac{5}{36I_1[\bar \nu_0-\varepsilon,\bar \nu_0,0]}+\epsilon_{\text{Pl}}\frac{5I_{1,\epsilon_{\text{Pl}}}[\bar \nu_0-\varepsilon,\bar\nu_0,0]}{36I_1[\bar \nu_0-\varepsilon,\bar \nu_0,0]^2 }.\nb\\
\eqn

Now let us turn to consider the second part of the error control function. Let us first define
\bqn
&&I_2[0,\bar y_0-\varepsilon,\epsilon_{\text{Pl}}]\nb\\
&&~~~~\equiv -\int_{\bar y_0-\varepsilon}^{0^+} \left\{\frac{q}{g}-\frac{5g'^2}{16g^3}+\frac{g''}{4g^2}\right\}\sqrt{g}dy'\nb\\
&&~~~~=\int_{\bar y_0-\varepsilon}^{0^+} G(y')dy',
\eqn
with
\bqn
G(y)\equiv-\left\{\frac{q}{g}-\frac{5g'^2}{16g^3}+\frac{g''}{4g^2}\right\}\sqrt{g}.
\eqn
According to the formula (\ref{derint}) we find
\bqn\lb{B14}
I_2[0,\bar y_0-\varepsilon,\epsilon_{\text{Pl}}] &\simeq& \int_{\bar \nu_0-\varepsilon}^{0^+} G(y')|_{\epsilon_{\text{Pl}}=0}dy'\nb\\
&&+\epsilon_{\text{Pl}} \int_{\bar \nu_0-\varepsilon}^{0^+} G_{,\epsilon_{\text{Pl}}}(y')|_{\epsilon_{\text{Pl}}=0}dy'\nb\\
&&-\epsilon_{\text{Pl}}\bar y_{0,\epsilon_{\text{Pl}}} G(\bar \nu_0-\varepsilon)|_{\epsilon_{\text{Pl}}=0}.
\eqn
Thus finally we can calculate the error control function by using the following formulas
\bqn\lb{int_exp}
\mathscr{H}(+\infty) &\simeq& -\frac{5}{36I_1[\bar \nu_0-\varepsilon,\bar \nu_0,0]}+\epsilon_{\text{Pl}}\frac{5I_{1,\epsilon_{\text{Pl}}}[\bar \nu_0-\varepsilon,\bar\nu_0,0]}{36I_1[\bar \nu_0-\varepsilon,\bar \nu_0,0]^2 }\nb\\&&+I_2[0,\bar y_0-\varepsilon,\epsilon_{\text{Pl}}],
\eqn
where $I_1[\bar \nu_0-\varepsilon,\bar \nu_0,0]$, $I_{1,\epsilon_{\text{Pl}}}[\bar \nu_0-\varepsilon,\bar\nu_0,0]$, and $I_2[0,\bar y_0-\varepsilon,\epsilon_{\text{Pl}}]$ are given by Eq.(\ref{B5}), Eq.(\ref{B9}), and Eq.(\ref{B14}), respectively.

\section*{Appendix C:  Functions $\mathscr{G}_i$ and  $\mathcal{\bar Q}_{j}^{(s)}$}
\renewcommand{\theequation}{C.\arabic{equation}} \setcounter{equation}{0}

The function $\mathscr{G}_{1,2}$ appearing in Eq.(\ref{sl_sqrt_g}) are given by
\bqn
\lb{C.1}
\mathscr{G}_1 &\equiv& \frac{\sqrt{\pi} \Gamma\left(\frac{\sigma}{2}\right)}{8\Gamma\left(\frac{\sigma+3}{2}\right)},\nb\\
\mathscr{G}_2 &\equiv& \frac{ \sqrt{\pi }  \Gamma \left(\frac{\sigma }{2}\right) }{8 \Gamma \left(\frac{\sigma +1}{2}\right)}\left[ \psi ^{(0)}\left(\frac{\sigma }{2}\right)-\psi ^{(0)}\left(\frac{\sigma +1}{2}\right)\right].
\eqn
Here $\psi^{(0)}(x)$ denotes the PolyGamma function. 
The function $\mathscr{G}_{3, 4, 5}$ appearing in Eq.(\ref{sl_sqrt_ga}) are given by
\bqn
\lb{C.2}
\mathscr{G}_3 & \equiv& -\frac{\sqrt{\pi } \Gamma \left(1+\frac{\sigma }{2}\right)  \left(\sigma^2-3\sigma+2\right)}{24 \Gamma \left(\frac{1+\sigma }{2}\right)},\nb\\
\mathscr{G}_4 & \equiv& \frac{\sqrt{\pi } (1+\sigma ) \Gamma \left(1+\frac{\sigma }{2}\right)\left(2 \sigma +\sigma ^2\right) }{48 \Gamma \left(\frac{3+\sigma }{2}\right)},\nb\\
\mathscr{G}_5 &\equiv& \frac{\sqrt{\pi } \Gamma \left(\frac{\sigma }{2}\right) }{48\sigma  (\sigma -3)  \Gamma \left(\frac{\sigma-1}{2}\right)}\nb\\
&&\times\Bigg\{\left(6 \sigma -5 \sigma ^2+\sigma ^3\right) \left[\psi ^{(0)}\left(\frac{\sigma-3}{2}\right)-\psi ^{(0)}\left(\frac{\sigma }{2}\right)\right]\nb\\
&&~~~~-\left(2 \sigma ^2-12 \sigma +12\right)\Bigg\}.
\eqn

The function $\mathcal{\bar Q}_{k}^{(s)}$ appearing in Eq.(\ref{scalar_spectrum}) are given by

\bqn
\lb{C.3}
\mathcal{\bar Q}_{-1}^{(s)}&=&-\frac{\sigma ^2 \left(\sigma ^2-2 \sigma -3\right)}{3}\alpha _0 \mathscr{G}_1 +\frac{40 \sigma ^2 (\sigma -3)}{543} \alpha _0\mathscr{G}_3 ,\nb\\
\mathcal{\bar Q}_{0}^{(s)} &=& \left(\sigma ^3-\frac{\sigma ^4}{3}\right)\alpha _0 \mathscr{G}_2+ \left(\frac{40 \sigma ^3}{181}-\frac{40 \sigma ^4}{543}\right)\alpha _0 \mathscr{G}_5\nb\\
&&+ \left(\frac{80 \sigma ^2}{181}-\frac{80 \sigma ^3}{543}\right) \alpha _0 \mathscr{G}_3\ln2\nb\\
&&+ \left(\frac{40 \sigma ^4}{1629}+\frac{208 \sigma ^3}{1629}-\frac{368 \sigma ^2}{543}\right)\alpha _0 \mathscr{G}_3\nb\\
&&+ \left(-\frac{2 \sigma ^4}{3}+\frac{4 \sigma ^3}{3}+2 \sigma ^2\right) \alpha _0 \mathscr{G}_1\ln2\nb\\
&&+ \left(\frac{\sigma ^5}{9}+\frac{134 \sigma ^4}{543}-\frac{2614 \sigma ^3}{1629}-\frac{315 \sigma ^2}{181}\right)\alpha _0 \mathscr{G}_1\nb\\
&&+\left(-\frac{40 \sigma ^3}{1629}-\frac{40 \sigma ^2}{543}+\frac{80 \sigma }{181}\right)\vartheta _0\mathscr{G}_3 \nb\\
&&+ \left(-\frac{\sigma ^4}{9}-\frac{4 \sigma ^3}{9}+\frac{5 \sigma ^2}{3}+2 \sigma \right)\vartheta _0 \mathscr{G}_1\nb\\
&& -3  \sigma  \chi \mathscr{G}_1+\frac{120  \chi }{181}\mathscr{G}_4,\nb\\
\mathcal{\bar Q}_{1}^{(s)}&=& \left(\sigma ^2-\frac{\sigma ^3}{3}\right)\alpha _0 \mathscr{G}_2+ \left(\frac{40 \sigma ^2}{181}-\frac{40 \sigma ^3}{543}\right)\alpha _0 \mathscr{G}_5\nb\\
&&+ \left(\frac{40 \sigma ^2}{181}-\frac{40 \sigma ^3}{543}\right) \alpha _0 \mathscr{G}_3\ln2\nb\\
&&+ \left(-\frac{40 \sigma ^4}{1629}+\frac{284 \sigma ^3}{1629}-\frac{68 \sigma ^2}{181}+\frac{40 \sigma }{181}\right)\alpha _0 \mathscr{G}_3\nb\\
&&+ \left(-\frac{\sigma ^4}{3}+\frac{2 \sigma ^3}{3}+\sigma ^2\right)\alpha _0 \mathscr{G}_1 \ln2\nb\\
&&+ \left(-\frac{\sigma ^5}{9}+\frac{248 \sigma ^4}{543}-\frac{764 \sigma ^3}{1629}-\frac{20 \sigma ^2}{543}+\sigma \right)\alpha _0 \mathscr{G}_1.\nb\\
\eqn

\end{document}